\documentclass[12pt]{article}

\usepackage{graphicx}

\usepackage{float}

\usepackage{xcolor}

\usepackage{amssymb}

\usepackage{lineno}

\usepackage{threeparttable}

\begin{document}

\title{Skewed non-Gaussian GARCH models for cryptocurrencies volatility modelling}

\author{Roy Cerqueti\\Sapienza University of Rome, Department of Economics and Law\\ London South Bank University, School of Business
\and Massimiliano Giacalone\footnote{Corresponding author, email: massimiliano.giacalone@unina.it}, Raffaele Mattera\\University of Naples "Federico II", Department of Economics and Statistics}

\date{}

\maketitle

\begin{abstract}
Recently, cryptocurrencies have attracted a growing interest from investors, practitioners and researchers. Nevertheless, few studies have focused on the predictability of them. In this paper we propose a new and comprehensive study about cryptocurrency market, evaluating the forecasting performance for three of the most important cryptocurrencies (Bitcoin, Ethereum and Litecoin) in terms of market capitalization. At this aim, we consider non-Gaussian GARCH volatility models, which form a class of stochastic recursive systems commonly adopted for financial predictions. Results show that the best specification and forecasting accuracy are achieved under the Skewed Generalized Error Distribution when Bitcoin/USD and Litecoin/USD exchange rates are considered, while the best performances are obtained for skewed Distribution in the case of Ethereum/USD exchange rate. The obtain findings state the effectiveness -- in terms of prediction performance -- of relaxing the normality assumption and considering skewed distributions.
\end{abstract}

\textbf{Keywords}: Generalized Error Distribution, GARCH models, Skewed distributions, volatility forecasting, non linear GARCH


\section{Introduction}
\label{S:1}

In the context of financial markets, an important problem is to define useful and efficient statistical methods for estimating and forecasting returns volatility. Indeed, the volatility of assets returns contributes to describe the riskiness of portfolios of assets, and its monitoring is thus of paramount relevance for management purposes \cite{stehlik2017}.

The volatility of a risky asset is strongly related to the way in which asset return evolves. In this respect, it is important to properly model the randomness of asset returns. The starting point of a good modeling exercise is unavoidably the observation of the empirical series of the returns \cite{sriboonchitta2013}.

As suggested by several authors (e.g. \cite{cont2001}), the time series of asset returns show very peculiar characteristics, since usually their distribution is asymmetric, with heavy-tails and negative skewness (\cite{chu1996stock}, \cite{durante2015}). Other empirical stylized facts on asset returns are also the presence of the so-called volatility clustering, conditional heteroskedasticity and the long-term memory property. (e.g. \cite{almeida2014}, \cite{wan2017})

For all these reasons, the Normal distribution is not a reliable choice for volatility modeling purposes,
and more sophisticated probabilistic assumptions which accounts, among the others, for normality deviation are needed (see e.g. \cite{orra2010} and references therein contained).

Such an observation offers a visualization of the volatility as a complex systems. For this reason, the analysis of such a key financial quantity and the assessment of methods for forecasting it are at the center of the debate of a large set of information scientists (see e.g. \cite{anzilli2018} and \cite{capotorti2013})

This paper contributes to the debate on volatility forecasting under non-Normal hypothesis for assets returns.
The proposed volatility forecasting methodology is based on Generalized Autoregressive Conditionally Heteroskedastic (GARCH) models, introduced by \cite{bollerslev1986} as a natural generalization of the ARCH models of \cite{engle1982}.

The GARCH model is of particular effectiveness for our purposes, since it is a stochastic system widely used for modelling the properties of randomness and uncertainity which chatacterize the volatility of the financial assets returns. Even if the original GARCH framework has been presented as a Gaussian-driven model, such a system allows for different kind of specifications to be adapted to modelling purposes (see e.g. \cite{almeida2014, hung2009fuzzy, efendi2018new}).

Accordingly to the arguments above, we depart from the standard Normal assumption and consider GARCH models under non-Gaussian distributional assumption.

In so doing, we are in line with a wide strand of literature, mainly for time series description or volatility estimation (see e.g. \cite{benos1997alternative} and \cite{durso16}). We mention also the $t$-student distribution approach of \cite{alberg2008}, which allows for a clear description of heavy tails characteristic of asset returns.

We propose a deep analysis of the volatility forecasting under non-Normal specifications for the resulting GARCH model by pointing our attention to the paradigmatic empirical case of cryptocurrencies, since several studies have observed that these types of assets are very highly volatile (see e.g. \cite{bariviera2017, baek2015}).

In particular, we here aim at identifying a probability distribution to model GARCH-based volatility for obtaining accurate forecasts. To pursue this scope, we provide a detailed analysis of the forecasting performances by employing the Generalized Error Distribution (GED) and its skewed version as distributional assumption. In particular, we empirically show that such a distributional assumption represents a suitable choice for volatility prediction purposes. In so doing, we offer also a confirmation of its flexibility (see e.g. the review in \cite{chenIS18}).

Cryptocurrencies are relatively a new type of asset (see e.g. \cite{gonzalez2019}) and the literature on this field is rapidly growing, even if it is still not well developed. Blockchain is the core technology employed for the creation of the cryptocurrencies. 
Such a technological device acts through the maintenance of immutable distributed ledgers in thousands of nodes. Thanks to the transactions' trustworthiness in the blockchain network, new cryptocurrencies are appearing in the financial markets \cite{chen2018exploring}.

One of the most popular members of the family of cryptocurrencies is the Bitcoin. Indeed, Bitcoin has a market capitalization higher than the one of the other cryptocurrencies (as Ethereum, Ripple, Litecoin, etc.). Despite such a predominance, one can observe an increasing competition among cryptocurrencies. Indeed, the Bitcoin’s market share fell down from the 80\% in the end of May 2016 to 48\% in the end of May 2017; in 2020, the Bitcoin's share is around 38\% (information available on 
coinmarketcap.com/charts).


In the empirical analysis, we show that the skewed specifications of the GARCH model represents the most effective selection for volatility forecasting of the Bitcoin/USD, Litecoin/USD and Ethereum/USD exchange rates, with a predominance of the GED distribution in the peculiar cases of Bitcoin and Litecoin.

Such results go in the direction of confirming the above mentioned stylized facts on the volatility of the cryptocurrencies. Findings have been validated by using a wide set of comparison loss functions and a wide set of alternative models. Some robustness checks have been also presented, to further support the main outcomes of the study.

The paper is structured as follows. Section \ref{sec2} contains a discussion on the employment of the Generalized Error Distribution (GED) in forecasting volatility under GARCH modeling. In Section \ref{sec3}, we provide a literature review on relevant previous studies related to cryptocurrencies volatility modelling. Section \ref{secdata} is devoted to the description of the considered empirical dataset, along with the methodologies used to analyze it. Section \ref{secempi} provides the illustration and the discussion of the obtained results. Section \ref{secrobu} presents the robustness check, which further supports the worthiness of the obtained empirical findings. Last Section offers some conclusive remarks and traces lines for future research.

\section{GARCH modeling with Generalized Error Distribution}
\label{sec2}

The volatility of assets returns is a crucial financial quantity, whose usefulness can be appreciated in a number of contexts like asset allocation, option pricing and risk management. The efficient estimation and prediction of the volatility is then of particular relevance, to gain insights about the future dynamics of prices and returns. Initially, assuming the general framework in which the Normal distribution assumption is not violated, methodological devices to estimate and forecast the volatility have been based on ARCH \cite{engle1982} and GARCH \cite{bollerslev1986} models- ARCH and GARCH are based on conditional heteroskedasticity of asset returns volatility. 

As already mentioned above, we here propose a new version of the GARCH models in the context of non-Normal distributions.

Given two integers $p,q >0$, we formalize the GARCH(p,q) model for the volatility $(\sigma^2_t: t \geq 0)$ as:

\begin{equation}
\label{garch}
\sigma^2_t = \omega + \sum_{i=1}^{p}\alpha_i z^2_{t-i}  + \sum_{j=1}^{q}\beta_j \sigma^2_{t-j}
\end{equation}

with $\omega>0$ and $\alpha_i>0,\beta_j>0$, for each $i=1, \dots, p$ and $j=1, \dots, q$.

The positivity condition on $\omega$, the $\alpha$'a and the $\beta$'s ensures the positivity of the variance. The term $(z_t:t \geq 0)$ is a stochastic process with i.i.d. time-realizations, which is here assumed to follow a Generalized Error Distribution (GED). Such an assumption -- which departs from the standard Normal hypothesis of \cite{bollerslev1986} -- is a suitable choice due to its strong flexibility for modeling asset returns volatility dynamics. Indeed, as already argued in the Introduction, the normality assumption is too restrictive and not reliable if the aim is to model financial asset returns, which clearly show empirically a non-Gaussian distribution.

The GED (also called Exponential Power Function) random variable $X$ has the following probability density function (see e.g. \cite{giacalone2018} and references therein contained):

\begin{equation}
\label{fGED}
f(z;\mu_p, \sigma_p, p) = \frac{p{\rm exp}(\frac{1}{2}|\frac{z-\mu_p}{\sigma_p}|^p)}{2p^{({1+\frac{1}{p})}}\sigma_p\Gamma(\frac{1}{p})}
\end{equation}

where $z \in \mathbb{R}$, $\mu_p \in (-\infty, +\infty)$ is called location parameter, $\sigma_p>0$  is called scale parameter, $p>0$ is a measure of fatness of tails and is called shape parameter and

\begin{equation}
\label{gamma}
\Gamma(a)=\int_{0}^{\infty}x^{a-1}{\rm exp}(-x)dx.
\end{equation}

Since the GED density function in (\ref{fGED}) is symmetric and unimodal, the location parameter is also the mode, median and mean of the distribution. The variance and kurtosis of the GED random variable are respectively given by:

$$
Var(X)=\sigma^2_p 2^\frac{2}{p} \frac{\Gamma(3p^{-1})}{\Gamma(p^{-1})}
$$
and
$$
Ku(X)=\frac{\Gamma(5p^{-1})}{\Gamma(3p^{-1})} \frac{\Gamma(p^{-1})}{\Gamma(3p^{-1})}.
$$

A very important feature of this family of distributions is that they include also other common distributions, for different values of shape parameter p. In particular, when $p=1$ we have a Laplace distribution, when $p=2$ we have the Gaussian distribution and for $p=+\infty$ we have the Uniform distribution. Moreover, the distribution has fatter tails than a Gaussian distribution when $p<2$ (see e.g. \cite{cerqueti2019generalized} and references therein contained).

\begin{figure}[H]
    \centering
    \includegraphics{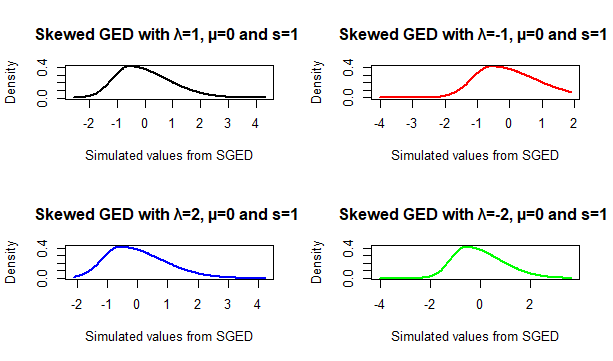}
    \caption{Skewed Generalized Error Distribution for different values of skewness.}
\label{Fig2}
\end{figure}

However, empirical evidence suggests that financial returns exhibit a negative symmetry in distribution; thus, we here propose to use skewed distribution in GARCH modeling (see \cite{theodossiou2015b}).
In this respect, we can hypothetically use either the Skew Normal or the Skew $t$ distributions. Nevertheless, according to the discussion above, a very interesting extension for skewness is the Skewed-GED distribution, which can be derived in order to take into account for the skewness and leptokurtosis (see Figure \ref{Fig2}).

The probability density function for non-centered Skewed GED can be defined as follow \cite{theodossiou2015b}:

\begin{equation}
f(z; \mu_p, \sigma_p, \lambda_p, p) = \frac{p{\rm exp}(\, -\frac{1}{p}|\frac{z-\mu_p + m}{\nu\sigma_p(1+\lambda_p {\rm sign}(\, z-\mu_p + m))}|^p)}{2\nu\sigma_p\Gamma(\, \frac{1}{p})}
\end{equation}

where $z \in \mathbb{R}$, $\mu_p$ is the location parameter, $\sigma_p$ is the scale parameter, $\lambda_p$ is the skewness parameter, $p$ is the shape parameter, while $\Gamma$ is as in (\ref{gamma}). Function ${\rm sign}$ is the sign function which assumes value of -1 for negative values of its argument and 1 for positive ones. Moreover, $m$ is defined as follow:
$$
m = \frac{2^{\frac{2}{p}}\nu\sigma_p\lambda_p\Gamma(\frac{1}{2}+\frac{1}{p})}{{\sqrt{\pi}}},
$$
while $\nu$:
$$
\nu = \frac{\pi(\, 1+3\lambda^2_p)\Gamma( \frac{3}{p})-16^{\frac{1}{p}}\lambda^2_p\Gamma(\frac{1}{2}+\frac{1}{p})\Gamma(\frac{1}{p})}{\pi\Gamma(\frac{1}{p})}.
$$

The shape parameter $p$ controls the tails and the peak of the distribution; a small value of $p$ means that the tails of the distribution become flat, with the center becoming largely peaked.

The skewness parameter $\lambda_p$ ranges in $[-1,1]$; in the case of negative skewness ($\lambda_p < 0$) the density function is skewed to the left and vice versa for $\lambda_p > 0$.

Also the Skewed GED (SGED) is a very special case of other distributions. For example, supposing $\lambda_p$=0 (allowing $p$ to change) we can obtain a wide family of non-skewed distributions.

In particular, when $\lambda_p= 0$ we have the GED; $\lambda_p = 0$ and $p=2$ means Normal distribution; $\lambda_p = 0$ and $p=\infty$ is the Uniform distribution and $\lambda_p=2$ and $p=2$ is the skewed Normal.

Also for the SGED-GARCH model the specification is the same as in (\ref{garch}), but in this case we suppose that $z_t$ follows a Skewed GED. The parameter estimation for the GED-GARCH models is based on the Maximum Likelihood method (see e.g. \cite{wisniewska2017}).

We will explore below the empirical effectiveness of the GED and its extension for skewness when predicting volatility through GARCH models.

Some further noticeable extensions of the GARCH models in (\ref{garch}) have been proposed in the literature, to remove the symmetry assumption in modeling volatility. We now provide a discussion on them. 

In \cite{glosten1993relation}, the so called GJR-GARCH model has been introduced as follows:

\begin{equation}
\label{gjrgarch}
    \sigma^2_t = \omega + \alpha z^2_{t-1}  + \beta \sigma^2_{t-1} + \gamma z^2_{t-1} I (z_{t-1} < 0),
\end{equation}

where $I(z_{t-1} < 0)$ which assigns 1 when $z_{t-1} < 0$ and 0 otherwise. If $\gamma = 0$, then (\ref{gjrgarch}) becomes (\ref{garch}) for $p=q=1$, so that we fall in the standard GARCH(1,1) case.

It is also worth mentioning the EGARCH model of \cite{nelson1991conditional} and the TGARCH model of \cite{zakoian1994threshold}. The main difference between TGARCH and GJR-GARCH -- that are quite similar for the rest -- is that TGARCH provides a modelization of the conditional standard deviation instead of the conditional variance.

Moreover, the classical GARCH model as in (\ref{garch}) can be also extended by accounting for highly persistence in conditional variances. Indeed, in the standard GARCH setting we know that one needs $\alpha + \beta < 1 $ -- i.e. the persistence of the conditional variance process is less than one -- in order to get that the unconditional variance $\sigma^2$ exists.

In this respect, equation (\ref{garch}) suggests that the presence of persistence is associated to a value of $\alpha + \beta$ close to the unity. Therefore, by imposing the restriction that $\alpha + \beta = 1$ in (\ref{garch}), we obtain the Integrated GARCH (IGARCH) model by analogy with the unit root literature:

\begin{equation}
    \sigma^2_t = \omega + \alpha (z^2_{t-1} - \sigma^2_{t-1}) + \sigma^2_{t-1}.
\end{equation}

Finally, another important extension -- which is mainly related to non linearity in terms of the parameters -- is the Asymmetric Power Generalized Autoregressive Conditional Heteroskedasticity (APGARCH) proposed by \cite{ding1993long}:

\begin{equation}
    \sigma^{\delta}_t = \omega + \alpha (|z_{t-1}| - \lambda z_{t-1})^{2\delta} + \beta \sigma^{\delta}_{t-1}
\end{equation}

where $\delta > 0 $, $\omega > 0$, $\alpha > 0$, $\beta \geq 0$ and $| \lambda | \leq 1$. This is a very general model and includes, for example, the Asimmetric GARCH (AGARCH) by \cite{meitz2011parameter} by setting $\delta =1 $.

Obviously, all the mentioned models can be estimated under Generalized Error Distribution assumption. Thus, they are part of the GED-GARCH models family.

\section{Volatility models for cryptocurrencies: a review}
\label{sec3}

Bitcoin has attracted the interest of many investors, practitioners and researchers since its creation in 2008. From there on, also other cryptocurrencies raised over the market attracting an increasing interest in both practitioners and academicians.

Bitcoin’s daily volatility has been studied in several papers. However, most of the existing studies have focused on in-sample analysis, and the comparisons of the volatility models have been implemented only on the ground of information criteria.

A very important literature contribution on the comparison between GARCH models in terms of in-sample performance for Bitcoin data is \cite{katsiampa2017}. The author compares several AR(1)-GARCH models through predefined information criteria, and shows that the one with the best performance is the AR(1)-Component-GARCH(1,1). The study exhibits some limitations.
First, \cite{katsiampa2017} evaluates only Bitcoin data, without considering also other cryptocurrencies; second, it imposes an AR(1) structure for the mean component of the GARCH model without assessing for forecasting out-of-sample performances; third, it considers only Gaussian distribution, even showing non-normality of the data.

Another relevant paper is \cite{mattera2018}, where the authors find out that, among six alternative distributional assumptions, the best model fitting the Bitcoin data is the AR(1)-AP-ARCH based on $t$-student distribution. As for the limitations of the quoted paper, \cite{mattera2018} considers only Bitcoin data; moreover, it forces, again, an AR(1) structure for the mean equation of the models.
Also the quoted paper does not assess for out-of-sample forecasting performance; rather than this, it obtains similar results for in-sample evaluation between $t$-student and GED assumption in standard GARCH setting.

Other studies have focused on the volatility dynamics of the Bitcoin returns. In particular, \cite{charles2019volatility} provide some further evidence starting from \cite{katsiampa2017}. However, the authors still considered just Gaussian distribution and did not provide an out-of-sample analysis.

In \cite{chu2017garch}, the authors find evidence of volatility clustering and show that, among several models, GARCH-type specifications provide the best in-sample performance. Using asymmetric GARCH models, \cite{bouri2017a}, \cite{katsiampa2017}, \cite{stavroyiannis2018value} and \cite{mattera2018} investigate the response of the conditional variance to past positive and negative shocks, finding evidence of the leverage effect.

The contribution \cite{chu2017garch} analyses Bitcoin and other cryptocurrencies using GARCH-type models with different error distributions, concluding that the best models for estimating the Bitcoin volatility are the I-GARCH and GJR-GARCH models with Gaussian distributions.
However, this study has two limitations: first of all, the proposed method forces an AR(1) process for the mean equation of the GARCH-type models; secondly, a real out-of-sample analysis in terms of forecasting accuracy is still missing.

In \cite{liu2017garch}, the authors compare the GARCH models by assuming the Normal Reciprocal Inverse Gaussian (NRIG) distribution and the Gaussian and Student-$t$ error distributions, and conclude that the GARCH model with Student-$t$ errors estimates the volatility better than the other ones. However, the quoted paper does not deal with the analysis of the performance of the skewed models.
Moreover, no analysis is implemented to compare the standard GARCH model (introduced in \cite{bollerslev1986}) with its extensions.

In \cite{naimy2018modelling}, the authors provide one of the first out-of-sample analysis. More precisely, they compare the one-step-ahead volatility forecasts estimated by GARCH and EGARCH models with Gaussian and the alternative $t$-student distribution.
The authors conclude that the EGARCH models present the best performances with respect to the two alternatives (EWMA and GARCH). Nevertheless, also here no attention is paid to skewed models, even if for Bitcoin -- as well as for other cryptocurrencies -- skewness is a well known stylized fact. Moreover, no details are provided about forecasting methodology as well as for predictive accuracy comparison.

Thus, although some first attempts in providing out-of-sample comparisons are available in the literature, most of them do not consider the skewness into the models. Moreover, the forecasting methodologies are sometimes presented without details and the predictive accuracy comparisons are not showed. In this sense, a comprehensive out-of-sample comparison seems to be still needed.

This paper is in line with the quoted contributions under the point of view of the scientific ground. However, it departs from them by trying to fix the mentioned limitations.

\section{Data and methodology}
\label{secdata}

\begin{figure}[H]
    \centering
    \includegraphics[width=\textwidth]{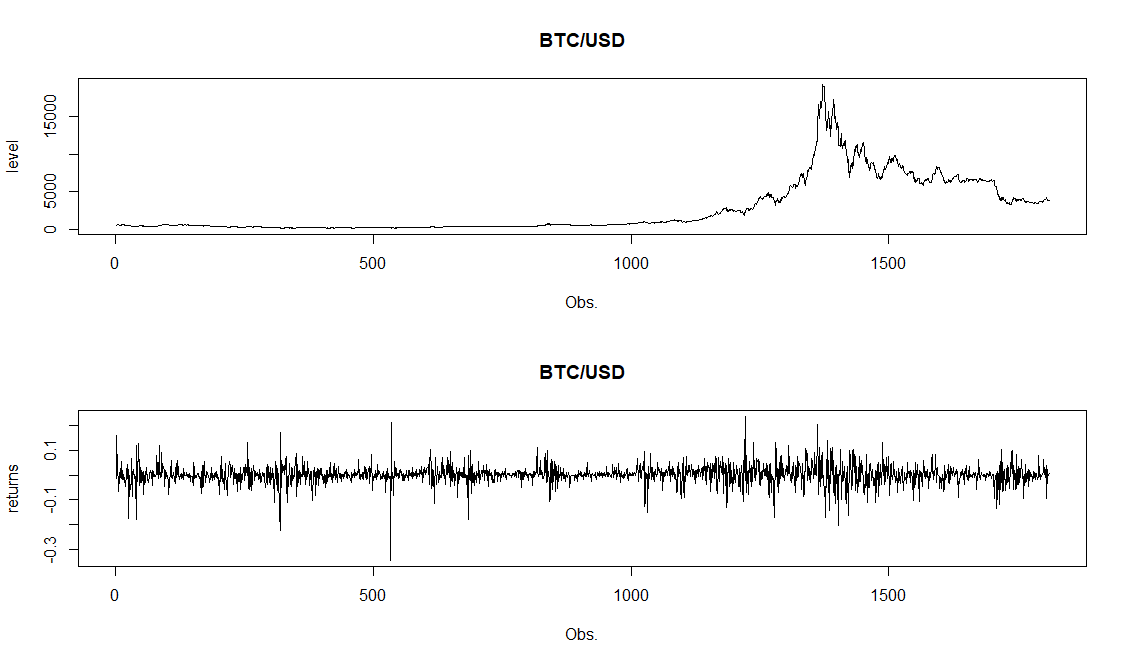}
    \caption{Bitcoin/US Dollar exchange rate versus its returns.}
\label{Fig2-1}
\end{figure}

The dataset contains the logarithm of last five-year daily exchange rates data (from March 2014 to March 2019) on the Bitfinex quotes for the most important cryptocurrencies: Bitcoin, Ethereum and Litecoin. In particular, we have selected the daily exchange rates with US Dollar (see Figures \ref{Fig2-1}, \ref{Fig4} and \ref{Fig6}), since such bilateral exchange rates are the most studied by previous literature due to data availability; moreover, they are also the most traded over the international stock markets. Data have been retrieved from the websites investing.com and www.bitfinex.com.

Since exchange rate time series are not stationary, we consider their returns as the ratio of the logarithm exchange rate values of two subsequent dates. We denote by $ER_t$ the logarithm of exchange rate value at time $t$. Then, the log-return $r_t$ between $t-1$ and $t$ is computed as follows:

$$
r_t = \frac{ER_t}{ER_{t-1}}
$$

The main descriptive statistics of the excahnge rates are showed in Table \ref{t1}.

\begin{table}[H]
\centering
\caption{Main descriptive statistics}
\begin{tabular}{l l l l l}
\hline
\multicolumn{5}{l}{Bitcoin/USD exchange rate} \\
\hline
\textbf{Mean} & \textbf{St. Dev.} & \textbf{Skewness} & \textbf{Kurtosis} & \textbf{Observations}\\
\hline
0.001065531 & 0.04023251 & -0.487245 & 7.45269 & 1811\\
\hline
\multicolumn{5}{l}{Ethereum/USD exchange rate} \\
\hline
\textbf{Mean} & \textbf{St. Dev.} & \textbf{Skewness} & \textbf{Kurtosis} & \textbf{Observations}\\
\hline
0.002270923 & 0.06223848 & -0.01642235 & 2.844286 & 1086\\
\hline
\multicolumn{5}{l}{Litecoin/USD exchange rate} \\
\hline
\textbf{Mean} & \textbf{St. Dev.} & \textbf{Skewness} & \textbf{Kurtosis} & \textbf{Observations}\\
\hline
0.002243284 & 0.05885434 & 1.50961 & 12.89685 & 1481\\
\hline
\end{tabular}
\label{t1}
\end{table}

However, the focus of the present study is related to the estimation of the parameters and to the volatility forecasting under Skewed non Gaussian models. In order to deal with our problem, a number of GARCH models for each exchange rate under several non Gaussian and Skewed distributions are proposed (see Table \ref{t2}).

\begin{table}[H]
\centering
\caption{Overview on implemented GARCH(1,1) models and their extensions}
\begin{tabular}{l}
\hline
\textbf{Models}\\
\hline
GARCH \\
GJR-GARCH \\
Treshold GARCH (TGARCH) \\
Exponential GARCH (EGARCH)\\
Integrated GARCH (IGARCH)\\
Asymmetric Power ARCH (APARCH)\\
\hline
\end{tabular}
\label{t2}
\end{table}

Moreover, we consider also several GARCH-type specifications, accounting for asymmetry and non-linearity (Table \ref{t3}). Therefore, overall we compare for each exchange rate 36 models.

\begin{table}[H]
\centering
\caption{Overview on distributional assumptions for GARCH-type models}
\begin{tabular}{l}
\hline
\textbf{Distributional assumptions}\\
\hline
Normal distribution \\
t-student distribution \\
Generalized Error Distribution \\
Skew Normal distribution\\
Skew t-student distribution\\
Skew Generalized Error Distribution\\
\hline
\end{tabular}
\label{t3}
\end{table}

All the models are fitted as being of GARCH(1,1) type, since in the practice this is the most convenient and parsimonous choice. This said, we also seek for the most appropriate selection of the GARCH model for the mean equation. In this direction, an automatic procedure involving several ARIMA models with the aim of selecting the one with the lowest Akaike Information Criterion (Table \ref{t4}) for all the considered cryptocurrencies has been implemented.

\begin{table}[H]
\centering
\caption{Results from mean equation process}
\begin{tabular}{l l}
\hline
\textbf{Exchange rate} & {ARIMA model}\\
\hline
BTC/USD & AR(2) \\
ETH/USD & ARMA(4,3) \\
LTC/USD & ARMA(2,2) \\
\hline
\end{tabular}
\label{t4}
\end{table}

Then, to evaluate which model gives in general a better specification in terms of goodness of fit and information, we consider the Akaike Information Criterion (AIC), that is one of the most used criteria at this aim (see e.g. \cite{wilhelmsson2006}).

In the end, the goodness of the performance of the volatility forecast has been tested.

The approach used in the forecasting exercise is of rolling window type. In particular, for all the considered exchange rates, we split the dataset in training and testing sets. While in the training phase we fit the model, for the testing period we implement the forecasting procedure and compare its results with the actual available realizations through some loss functions' values. The testing set is composed of the last 200 observations of the dataset. In the rolling window approach, windows are shifted by one date.

As a preliminary step, we identify the loss functions to be used. Among them, we reasonably include  the Mean Square Error (MSE), which is the most popular one.
Moreover, \cite{patton2011volatility} found that the MSE is the most robust loss function when used to compare volatility forecasting models.

However, it is well-known that MSE can be possibly inflated by the presence of outliers; thus, we take into account also the Mean Absolute Error (MAE) \cite{amendola2002} and the Root Mean Square Error (RMSE) \cite{poon2003forecasting}.

To present more robust results, we have compared the predictive accuracy of the forecasts according to the above mentioned loss functions by using a statistical test. In particular, to serve this scope, we have implemented the test in \cite{diebold2002comparing}.
The \cite{diebold2002comparing} test assesses wheter the forecasts of two different models statistically differ in terms of predictive accuracy. Only when two models provide statistically different forecasts, we would be able to correctly disentangle what is the best one from the predictive point of view.
Hence, a brief presentation of the testing procedure is nedeed.

Consider two different statistical models $A$ and $B$. We can define the forecast errors as follows:
$$
e_{A,t} = \hat{y}_{A,t} - y_t
$$
and
$$
e_{B,t} = \hat{y}_{B,t} - y_t,
$$
where $\hat{y}_{A,t}$ and $\hat{y}_{B,t}$ are the predictions of models A and B, respectively, and $y_t$ is the actual observed value. Now, consider a generic loss function $g$ to be applied to the prediction error. The \cite{diebold2002comparing} procedure tests wheter the difference in forecasting accuracy is equal or different from zero. Formally, we define the difference in forecasting accuracy as:
$$
d_t = g(e_{A,t}) - g(e_{B,t}).
$$
Under the null hypotesis of equal predictive accuracy we have that:
$$
E(d_t) = 0,
$$
while under the alternative hypotesis we have:
$$
E(d_t) \neq 0.
$$
It is worth mentioning that the test statistics follow a standard normal distribution under the null hypotesis.

\section{Empirical experiments}
\label{secempi}

We here presents a validation of the theoretical setting, by dealing with some empirical exercises. As preannounced above, we propose the study of the exchange rates of three among the most popular cryptocurrencies -- i.e., three of the ones with the highest market capitalizations: Bitcoin, Ethereum and Litecoin -- with the USD -- which represents a worldwide acknowledge reference currency. This said, the empirical sample seems to be particularly representative of the exchange rates of cryptocurrencies with physical ones. The selection of the cryptocurrencies is based not only on their relevance in terms of market capitalization, but also on data availability. 

The results of the investigations are presented by distinguishing the different cryptocurrencies, for the sake of clarity.

\subsection{Bitcoin data}
\label{subbit}

The first experiment is conducted on the most important cryptocurrency in terms of market capitalization (https://coinmarketcap.com). In particular, we study the dynamics of the exchange rate with US Dollars.

In order to prove that the data are non-normally distributed, we have performed the Jarque-Bera test for normality. The result of the Jarque-Bera test is  4275.932 with a null p-value, which means that we can reject the null hypothesis that residuals follow a normal distribution. These results confirm the reason of the alternative distribution based GARCH model adoption instead of a “Gaussian” GARCH model.

So, by proceeding with the parameter estimation of the standard GARCH(1,1) model based on normality, we found the results collected in Table \ref{t5}.

\begin{table}[H]
\centering
\caption{Estimation from Gaussian GARCH(1,1) model}
\begin{tabular}{l l l}
\hline
\textbf{} & {Coefficient} & {Standard Error}\\
\hline
$\omega$ & 0.000058 & 0.000058 \\
$\alpha$ & 0.110905*** & 0.023857 \\
$\beta$ & 0.861032*** & 0.029985 \\
\hline
\end{tabular}
\begin{tablenotes}
      \small
      \item Note: *** means significance at 1\%, ** at 5\% and * at 10\%, standard errors are computed as robust.
    \end{tablenotes}
\label{t5}
\end{table}

\begin{figure}[H]
    \centering
    \includegraphics[width=\textwidth]{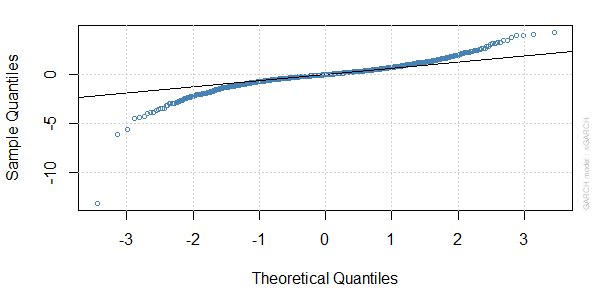}
    \caption{Q-Q plot of standardized residuals from Gaussian GARCH(1,1).}
\label{Fig3}
\end{figure}

\begin{figure}[H]
    \centering
    \includegraphics[width=\textwidth]{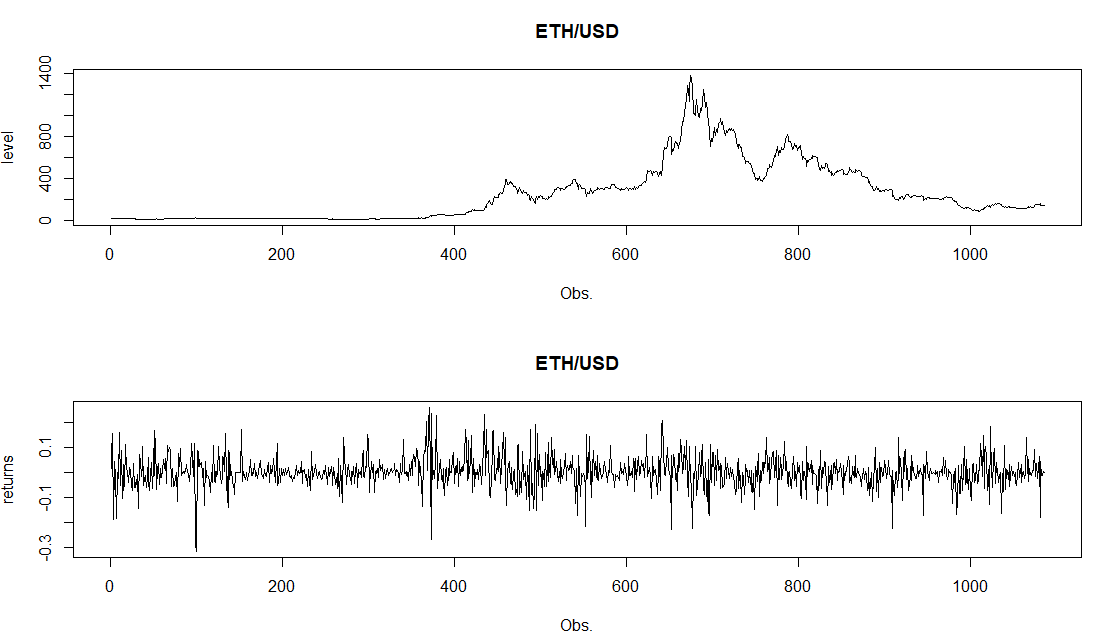}
    \caption{Ethereum/US Dollar exchange rate versus its returns.}
\label{Fig4}
\end{figure}

After the parameters’ estimation, we have analyzed also the Q-Q plot of standardized residuals to see if normality assumption holds for the specified model (Fig. \ref{Fig3}).

Considering the residuals shape in the plot, the normality assumption seems to be violated. This result give us an additional element to apply another distributional assumption in GARCH(1,1) model for the volatility analysis.

On the light of these results, we have estimated the parameters for all the alternative methods, founding that all parameters are significant and that the standard errors are smaller in the GED-based GARCH models than in the other alternative ones.

Indeed, among the alternative models, the one with lowest standard errors is the Skewed GED-GARCH. Results are showed in the Table \ref{t6}.

\begin{table}[H]
\centering
\caption{Results from the alternative GARCH(1,1) models}
\begin{tabular}{l l l l l l}
\hline
\textbf{} & {Skew Normal} & {t-student} & {Skew t-student} & {GED} & {Skew GED}\\
\hline
$\omega$ & 0.000056* & 0.000023* & 0.000023* & 0.000024** & 0.000024** \\
       & (0.000034) & (0.000015) & (0.000015) & (0.000012) & (0.000009)\\
$\alpha$ & 0.116420*** & 0.145805*** & 0.146566*** & 0.139268*** & 0.140590***\\
       & (0.023344) & (0.021306) & (0.021620) & (0.023730) & (0.020849)\\
$\beta$  & 0.857186*** & 0.853195*** & 0.852434*** & 0.859724*** & 0.858406***\\
       & (0.029306)	& (0.030470) & (0.030652) & (0.025555) & (0.019411)\\
\hline
\end{tabular}
\begin{tablenotes}
      \small
      \item Note: *** means significance at 1\%, ** at 5\% and * at 10\%, robust standard errors in parenthesis.
    \end{tablenotes}
\label{t6}
\end{table}

We have estimated parameters also for the other considered GARCH(1,1)-type as in Table \ref{t2}, finding the same results.
After the parameter estimation, we have assessed also for model specification (see Tables \ref{t7} and \ref{t8}). Indeed, following the AIC and BIC criteria, it is clear that we cannot obtain a good specification with a normality-based GARCH model.

In particular, it is clear that better results in terms of specification are obtained when considering skewed distributions. Moreover, relaxing the standard GARCH(1,1) specification allows us to obtain a better data fitting, since the lowest AIC and BIC values are associated to the Treshold GARCH(1,1).

Nevertheless, in order to assess for the best model, the forecasting performances have been also considered. The quality of the forecast is evaluated in Tables \ref{t9} and \ref{t10}.

\begin{table}[H]
\centering
\caption{Information criteria for all GARCH models}
\begin{tabular}{l l l}
\hline
\textbf {Distribution} & {AIC} & {BIC}\\
\hline
GARCH(1,1)\\
\hline
Normal & -3.7751 & -3.7567 \\
Skew Normal & -3.7919 & -3.7704 \\
t-student & -4.0834 & -4.0619 \\
Skew t-student & -4.0827 & -4.0582 \\
GED & -4.0796 & -4.0581 \\
Skew GED & -4.0787 & -4.0542 \\
\hline
GJR-GARCH(1,1)\\
\hline
Normal & -3.7471 & -3.7237 \\
Skew Normal & -3.7619 & -3.7351 \\
t-student & -4.0454 & -4.0187 \\
Skew t-student & -4.0446 & -4.0145 \\
GED & -4.0410 & -4.0142 \\
Skew GED & -4.0397 & -4.0542 \\
\hline
T-GARCH(1,1)\\
\hline
Normal & -3.7536 & -3.7302 \\
Skew Normal & -3.7600 & -3.7332 \\
t-student & -4.0630 & -4.0363 \\
Skew t-student & -4.0621 & -4.0320 \\
GED & -4.0821 & -4.0873 \\
Skew GED & -4.0844 & -4.0844 \\
\hline
\end{tabular}
\label{t7}
\begin{tablenotes}
      \small
      \item Note: AIC and BIC are Akaike Information Criterion and Bayesian Information Criterion, respectively. The lowest value is associated to the best fitting.
\end{tablenotes}
\end{table}

\begin{table}[H]
\centering
\caption{Information criteria for all GARCH models}
\begin{tabular}{l l l}
\hline
\textbf {Distribution} & {AIC} & {BIC}\\
\hline
E-GARCH(1,1)\\
\hline
Normal & -3.7714 & -3.7480 \\
Skew Normal & -3.7812 & -3.7545 \\
t-student & -4.0596 & -4.0328 \\
Skew t-student & -4.0586 & -4.0285 \\
GED & -4.0490 & -4.0223 \\
Skew GED & -4.0478 & -4.0177 \\
\hline
I-GARCH(1,1)\\
\hline
Normal & -3.7455 & -3.7288 \\
Skew Normal & -3.7619 & -3.7418 \\
t-student & -4.0477 & -4.0277 \\
Skew t-student & -4.0433 & -4.0233 \\
GED & -4.0490 & -4.0223 \\
Skew GED & -4.0421 & -4.0187 \\
\hline
AP-ARCH(1,1)\\
\hline
Normal & -3.7527 & -3.7260 \\
Skew Normal & -3.7622 & -3.7322 \\
t-student & -4.0580 & -4.0279 \\
Skew t-student & -4.0614 & -4.0279 \\
GED & -4.0477 & -4.0176 \\
Skew GED & -4.0477 & -4.0176 \\
\hline
\end{tabular}
\label{t8}
\begin{tablenotes}
      \small
      \item Note: AIC and BIC are Akaike Information Criterion and Bayesian Information Criterion, respectively. The lowest value is associated to the best fitting.
\end{tablenotes}
\end{table}

\begin{table}[H]
\centering
\caption{Volatility forecasting performance for GARCH(1,1)-type models}
\begin{tabular}{l l l l}
\hline
\textbf {Distribution} & {MSE} & {MAE} & {RMSE}\\
\hline
GARCH(1,1)\\
\hline
Normal$^{\dagger}$ & 0.00126238 & 0.03411553 &  0.03553010\\
Skew Normal & 0.00124281*** & 0.03379684*** & 0.03525351***\\
t-student & 0.00118389*** & 0.03217911*** & 0.03440784*** \\
Skew t-student & 0.00118247*** & 0.03215938*** & 0.03438717*** \\
GED & 0.00118126*** & 0.03220527*** & 0.03436953***\\
Skew GED & 0.00118058*** & 0.03219489*** & 0.03435968***\\
\hline
GJR-GARCH(1,1)\\
\hline
Normal & 0.00130525*** & 0.03455626*** & 0.03612829***\\
Skew Normal & 0.00125908 & 0.03397627 & 0.03548353 \\
t-student & 0.00115346*** & 0.03181912*** & 0.03396266***\\
Skew t-student & 0.00115211*** & 0.03179883*** & 0.03394274*** \\
GED &  0.00115846*** & 0.03194733*** &  0.03403623***\\
Skew GED & 0.00115839*** & 0.0.03194536*** & 0.03403522***\\
\hline
T-GARCH(1,1)\\
\hline
Normal & 0.00132715*** & 0.03446651*** &   0.03643012***\\
Skew Normal & 0.00128298 & 0.03393094 & 0.03581883\\
t-student & 0.00163671*** & 0.03723041*** & 0.04045636*** \\
Skew t-student & 0.00164403*** & 0.03730592*** & 0.04054673*** \\
GED & 0.00012554*** & 0.01008683*** & 0.01120463***\\
Skew GED & 0.00012554*** & 0.01008683*** &  0.01120463***\\
\hline
\end{tabular}
\label{t9}
\begin{tablenotes}
      \small
      \item Note: *** means significance at 1\%, ** at 5\% and * at 10\%, otherwise no significance for \cite{diebold2002comparing} test of predictive accuracy compared with GARCH(1,1) under normal distribution ($\dagger$ recognizes the benchmark model). Under the null we have equal predictive accuracy.
\end{tablenotes}
\end{table}

\begin{table}[H]
\centering
\caption{Volatility forecasting performance for GARCH(1,1)-type models}
\begin{tabular}{l l l l}
\hline
\textbf {Distribution} & {MSE} & {MAE} & {RMSE}\\
\hline
E-GARCH(1,1)\\
\hline
Normal & 0.00130713*** &  0.03439897*** & 0.03615426*** \\
Skew Normal & 0.00126637 & 0.03387139 & 0.03558617\\
t-student & 0.00158607*** & 0.03712476*** & 0.03982559***\\
Skew t-student & 0.00158640*** & 0.03712274*** & 0.03982965***\\
GED & 0.00116151*** & 0.03200013*** & 0.03408098***\\
Skew GED & 0.00116034*** & 0.03199197*** & 0.03406389***\\
\hline
I-GARCH(1,1)\\
\hline
Normal & 0.00134033*** & 0.03480408*** & 0.03661061*** \\
Skew Normal & 0.00132019*** & 0.03449638*** & 0.03633453*** \\
t-student & 0.00118959*** & 0.03224817*** & 0.03449057***\\
Skew t-student & 0.00118813*** & 0.03222796*** & 0.03446932***\\
GED & 0.00118639*** & 0.03226152*** &  0.03444404***\\
Skew GED &  0.00118593*** & 0.03225488*** & 0.03443746*** \\
\hline
AP-ARCH(1,1)\\
\hline
Normal & 0.00132362*** & 0.03449170*** & 0.03638171***\\
Skew Normal & 0.00126376 & 0.03392030 & 0.03554941\\
t-student & 0.00151351*** & 0.03535722*** & 0.03890392***\\
Skew t-student & 0.00163545*** & 0.03712063*** & 0.04044070***\\
GED & 0.00118879*** & 0.03199220*** & 0.03447897***\\
Skew GED & 0.001159*** & 0.031805*** & 0.057321***  \\
\hline
\end{tabular}
\label{t10}
\begin{tablenotes}
      \small
      \item Note: *** means significance at 1\%, ** at 5\% and * at 10\%, otherwise no significance for \cite{diebold2002comparing} test of predictive accuracy compared with GARCH(1,1) under normal distribution. Under the null we have equal predictive accuracy.
\end{tablenotes}
\end{table}

The predictive accuracy test of \cite{diebold2002comparing} is reported -- 
along with forecast errors -- for all the models against the selected benchmark, i.e. the Gaussian standard GARCH model.

In the evaluation step, most of the models not only differ from the standard Gaussian GARCH but also outperform it. These results confirm the previous findings of \cite{mattera2018}.

In the experiments, the model with the lowest value of its loss function is the Threshold GARCH model based on Skewed Generalized Error Distribution.

Moreover, we have also investigated the difference in predictive accuracy of the skewed models when compared with the not skewed ones. In particular, we focus our attention to the differences between $t$-student and GED based alternatives for all the GARCH-type models (see Table \ref{t11}), especially on the light of the findings of the previous tables.

Indeed, Skewed Normal models, for most of the alternative GARCH-type specifications, have the same predictive accuracy of the simple GARCH(1,1) based on Gaussian distribution. This is precisely the reason for which we do not consider them in this case.

\begin{table}[H]
\centering
\caption{Predictive accuracy test between skewed and not skewed models}
\begin{tabular}{l l l}
\hline
\textbf {Distributions} & {} & {}\\
\hline
\textit{GARCH(1,1)} & {Skewed t-student} & {Skewed GED}\\
\hline
t-student & 3.7346*** &  1.6782* \\
GED & -0.59961 & 6.4659***\\
\hline
\textit{GJR-GARCH(1,1)} & {Skewed t-student} & {Skewed GED}\\
\hline
t-student & 3.9013*** &  -3.4718*** \\
GED & 3.8543*** & 0.03387139\\
\hline
\textit{T-GARCH(1,1)} & {Skewed t-student} & {Skewed GED}\\
\hline
t-student & -8.2745*** &  17.352*** \\
GED & -17.327*** & 0 \\
\hline
\textit{E-GARCH(1,1)} & {Skewed t-student} & {Skewed GED} \\
\hline
t-student & -0.56007 &  3.9004*** \\
GED & -16.334*** & 16.284***\\
\hline
\textit{I-GARCH(1,1)} & {Skewed t-student} & {Skewed GED}\\
\hline
t-student & 3.8079*** &  1.9074* \\
GED & -0.88552 & 5.6451***\\
\hline
\textit{AP-ARCH(1,1)} & {Skewed t-student} & {Skewed GED}\\
\hline
t-student & -5.6859*** &  24.231*** \\
GED & -13.888*** & 23.58***\\
\hline
\end{tabular}
\label{t11}
\begin{tablenotes}
      \small
      \item Note: the reported values are associated to the results of \cite{diebold2002comparing} test statistic under MSE loss function. *** means significance at 1\%, ** at 5\% and * at 10\%, otherwise no significance. Under the null we have equal predictive accuracy.
\end{tablenotes}
\end{table}

According to results in Table \ref{t11}, in only one case we have equal predictive accuracy between the classical $t$-student and its skewed extension (in the case og GJR-GARCH specification) and for GED and Skewed GED (for E-GARCH). Nevertheless, in most of the other cases, we do not obtain a similar  predictive accuracy. Moreover, one can easily notice remarkable discrepancies among different distributional families (e.g. $t$-student versus Generalized Error Distribution).

So, overall, forecasts obtained with skewed distribution statistically differ from the ones obtained from the same GARCH-type models but under not skewed distributions. According to results in Tables \ref{t9} and \ref{t10}, it is clear that skewed models outperform not skewed ones for Bitcoin data; moreover, the most accurate forecasting method is the SGED-T-GARCH.


\subsection{Ethereum data}
\label{sec5.2}

The second experiment is conducted on the second cryptocurrency in terms of market capitalization (https://coinmarketcap.com). As in the first application, we study the dynamics of the exchange rate with US Dollars.

As the previous experiment, the first step of is to assess for the model’s specification and parameters estimation.

However, again, we have to show first that GARCH(1,1) models with alternative distributions are more effective in modeling than the simple GARCH(1,1), when the returns follow a Gaussian distribution.

Also in this case, data are non-normally distributed according to the Jarque-Bera test for normality. The resulting test statistic is 368.8993 with a p-value close to zero, which means that we can reject the null hypothesis that residuals follow a normal distribution.

These results allow us to specify an alternative distribution-based GARCH model instead of a “Gaussian” GARCH one.

So, by proceeding with the parameters estimation of the standard normal GARCH(1,1), we found the results reported in Table \ref{t12}.

\begin{table}[H]
\centering
\caption{Estimation for Gaussian GARCH(1,1) model}
\begin{tabular}{l l l}
\hline
\textbf{} & {Coefficient} & {Standard Error}\\
\hline
$\omega$ & 0.000350** & 0.000350** \\
$\alpha$ & 0.000350** & 0.039690 \\
$\beta$ & 0.767006*** & 0.062498 \\
\hline
\end{tabular}
\begin{tablenotes}
      \small
      \item Note: *** means significance at 1\%, ** at 5\% and * at 10\%,
standard errors are computed as robust.
    \end{tablenotes}
\label{t12}
\end{table}

\begin{figure}[H]
    \centering
    \includegraphics[width=\textwidth]{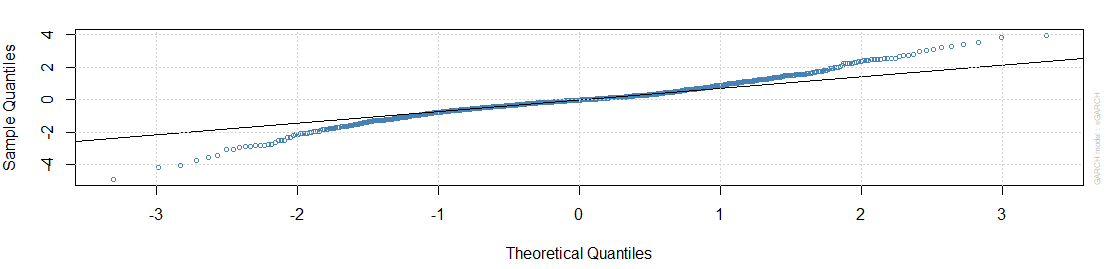}
    \caption{Q-Q plot of standardized residuals from Gaussian GARCH(1,1).}
\label{Fig5}
\end{figure}

\begin{figure}[H]
    \centering
    \includegraphics[width=\textwidth]{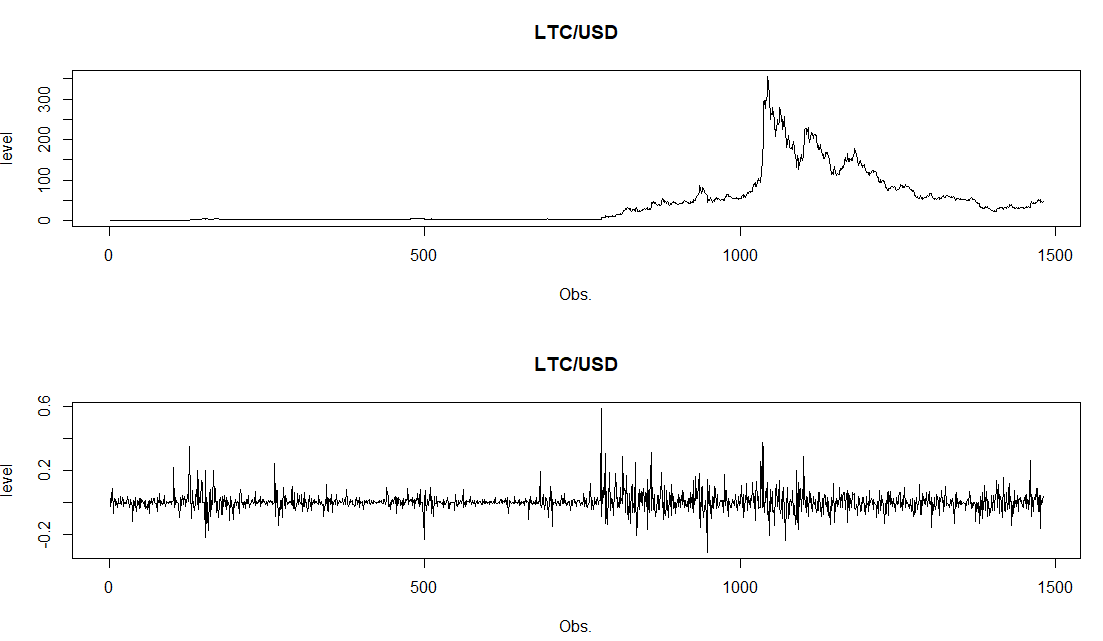}
    \caption{Litecoin/US Dollar exchange rate versus its returns.}
\label{Fig6}
\end{figure}

We have analyzed also the Q-Q plot of standardized residuals (Figure \ref{Fig5}). By considering the residuals shape in the plot, the normality assumption seems to be violated. This result gives us an additional element to employ a modification of the standard normal GARCH(1,1) model for the volatility analysis.

On the light of these results, we have estimated the parameters for all the alternative methods. Also from this second experiment, we found all significant parameters and smaller standard errors in the GED-based GARCH models than to the alternatives ones. The results are shown in the Table \ref{t13}.

\begin{table}[H]
\centering
\caption{Results from the alternative GARCH(1,1) models}
\begin{tabular}{l l l l l l}
\hline
\textbf{} & {Skew Normal} & {t-student} & {Skew t-student} & {GED} & {Skew GED}\\
\hline
$\omega$ & 0.000349*** & 0.000225** & 0.000221** & 0.000219* & 0.000211***\\
       & (0.000152)	& (0.000109) & (0.000104) & (0.000090) & (0.000043)\\
$\alpha$ & 0.140882*** & 0.203371*** & 	0.197759*** & 	0.169869*** & 0.164906***\\
       & (0.039668) & (0.045492) & (0.043893) & (0.042117) & (0.015247)\\
$\beta$  & 0.767360*** & 0.795629*** & 0. 801241*** & 0.794239*** & 0.802947***\\
       & (0.062097) & (0.044750) & (0.042540) & (0.043910) & (0.017428)\\
\hline
\end{tabular}
\begin{tablenotes}
      \small
      \item Note: *** means significance at 1\%, ** at 5\% and * at 10\%, robust standard errors in parenthesis.
    \end{tablenotes}
\label{t13}
\end{table}

After the parameters estimation, we have assessed also for model specification trough an in-sample analysis (see Tables \ref{t14} and \ref{t15}).

Indeed, following the AIC and BIC criteria, it is clear that a GARCH-type model based on normality fails in obtaining a good in-sample fitting. In particular, with the GED-GARCH(1,1) model, we obtain the smallest value and therefore the best fit.

This conclusion applies for all the alternative GARCH-type models, where almost always GED-based GARCH models provide the best in-sample performances. More precisely, the GED-based I-GARCH model is the best fitting one, even if GJR-GARCH and T-GARCH alternatives have close information criteria values.

\begin{table}[H]
\centering
\caption{Information criteria for all GARCH models}
\begin{tabular}{l l l}
\hline
\textbf {Distribution} & {AIC} & {BIC}\\
\hline
GARCH(1,1)\\
\hline
Normal &  -2.8333 & -2.7739 \\
Skew Normal & -2.8326 & -2.7678 \\
t-student & -2.9276 & -2.8628 \\
Skew t-student & -2.9275 & -2.8572 \\
GED & -2.9670 & -2.9022 \\
Skew GED & -2.9672 & -2.8970 \\
\hline
GJR-GARCH(1,1)\\
\hline
Normal & -2.8310 & -2.7662 \\
Skew Normal & -2.8304 & -2.7601 \\
t-student & -2.9259 & -2.8556 \\
Skew t-student & -2.9257 & -2.8501 \\
GED &  -2.9702 & -2.9000 \\
Skew GED & -2.9650 &  -2.8894 \\
\hline
T-GARCH(1,1)\\
\hline
Normal & -2.8125 & -2.7477 \\
Skew Normal & -2.8086 & -2.7383 \\
t-student & -2.9255 & -2.8552 \\
Skew t-student & -2.9255 & -2.8499 \\
GED & -2.9615 & -2.8912 \\
Skew GED & -2.9571 & -2.8815 \\
\hline
\end{tabular}
\label{t14}
\begin{tablenotes}
      \small
      \item Note: AIC and BIC are Akaike Information Criterion and Bayesian Information Criterion, respectively. The lowest value is associated to the best fitting.
\end{tablenotes}
\end{table}

\begin{table}[H]
\centering
\caption{Information criteria for all GARCH models}
\begin{tabular}{l l l}
\hline
\textbf {Distribution} & {AIC} & {BIC}\\
\hline
E-GARCH(1,1)\\
\hline
Normal & -2.8339 & -2.7691 \\
Skew Normal &  -2.8324 & -2.7622 \\
t-student & -2.9320 & -2.8618 \\
Skew t-student & -2.9319 & -2.8562 \\
GED & -2.9685 & -2.8983 \\
Skew GED & -2.9679 & -2.8923 \\
\hline
I-GARCH(1,1)\\
\hline
Normal & -2.8282 & -2.7741 \\
Skew Normal & -2.8275 & -2.7681 \\
t-student & -2.9299 & -2.8704 \\
Skew t-student & -2.9297 & -2.8649 \\
GED & -2.9680 &  -2.9086 \\
Skew GED & -2.9670 & -2.9022 \\
\hline
AP-ARCH(1,1)\\
\hline
Normal & -2.8287 & -2.7585 \\
Skew Normal & -2.8324 & -2.7568 \\
t-student & -2.9248 & -2.8492 \\
Skew t-student & -2.9248 & -2.8437 \\
GED & -2.9632 & -2.8875 \\
Skew GED & -2.9674 & -2.8863 \\
\hline
\end{tabular}
\label{t15}
\begin{tablenotes}
      \small
      \item Note: AIC and BIC are Akaike Information Criterion and Bayesian Information Criterion, respectively. The lowest value is associated to the best fitting.
\end{tablenotes}
\end{table}

However, in order to detect the best performing model, we consider also in this case the forecasting performances. The quality of the forecast is evaluated in Tables \ref{t16} and \ref{t17}.

\begin{table}[H]
\centering
\caption{Volatility forecasting performance for GARCH(1,1)-type models}
\begin{tabular}{l l l l}
\hline
\textbf {Distribution} & {MSE} & {MAE} & {RMSE}\\
\hline
GARCH(1,1)\\
\hline
Normal$^{\dagger}$ & 0.003045198 & 0.05388708 &  0.05518331\\
Skew Normal &  0.003042917* & 0.05384609 & 0.05516265\\
t-student & 0.003651810*** & 0.05863692*** & 0.06043021*** \\
Skew t-student & 0.003672101*** & 0.05885401*** & 0.06059786*** \\
GED & 0.003218927*** & 0.05517174*** &  0.05673559***\\
Skew GED & 0.003234109*** & 0.05532379*** & 0.05686923***\\
\hline
GJR-GARCH(1,1)\\
\hline
Normal & 0.003036255 & 0.05378802 & 0.05510222\\
Skew Normal & 0.003041529*** & 0.05383584*** & 0.05515006*** \\
t-student & 0.003676746*** & 0.05876707*** & 0.06063618***\\
Skew t-student & 0.003691336*** & 0.05893010*** & 0.06075637*** \\
GED &  0.003327975*** & 0.05595303*** &  0.05768861***\\
Skew GED & 0.003314663*** & 0.05585799*** & 0.05757311***\\
\hline
T-GARCH(1,1)\\
\hline
Normal & 0.002913599*** & 0.05257935*** &   0.05397777***\\
Skew Normal & 0.002896731*** & 0.05242472*** & 0.05382128***\\
t-student & 0.003653328*** & 0.05846574*** & 0.06044277*** \\
Skew t-student & 0.003675612*** & 0.05869657*** & 0.06062683*** \\
GED &  0.003101052 & 0.05400137 & 0.05568709\\
Skew GED & 0.003106676 & 0.05416085 & 0.05573757\\
\hline
\end{tabular}
\label{t16}
\begin{tablenotes}
      \small
      \item Note: *** means significance at 1\%, ** at 5\% and * at 10\%, otherwise no significance for \cite{diebold2002comparing} test of predictive accuracy compared with GARCH(1,1) under normal distribution ($\dagger$ recognizes benchmark model). Under the null we have equal predictive accuracy.
\end{tablenotes}
\end{table}

\begin{table}[H]
\centering
\caption{Volatility forecasting performance for GARCH(1,1)-type models}
\begin{tabular}{l l l l}
\hline
\textbf {Distribution} & {MSE} & {MAE} & {RMSE}\\
\hline
E-GARCH(1,1)\\
\hline
Normal & 0.002973416* & 0.05320498 & 0.05452904 \\
Skew Normal &  0.002954408* & 0.05298161 & 0.05435446\\
t-student & 0.003675423*** & 0.05876958*** & 0.06062527***\\
Skew t-student & 0.003674148*** & 0.05880266*** & 0.06061475***\\
GED & 0.003253827*** & 0.05546392*** & 0.05704232***\\
Skew GED & 0.003141111*** & 0.05452894*** & 0.05604561***\\
\hline
I-GARCH(1,1)\\
\hline
Normal & 0.003397301*** & 0.05626348*** & 0.05828637*** \\
Skew Normal & 0.003389157*** & 0.05619944*** & 0.05821647*** \\
t-student & 0.003663309*** & 0.05872274*** & 0.06052528***\\
Skew t-student & 0.003684441*** & 0.05894666*** & 0.06069960***\\
GED & 0.003472444*** & 0.05693766*** & 0.05892745***\\
Skew GED & 0.003474597*** & 0.05699587*** & 0.05894571*** \\
\hline
AP-ARCH(1,1)\\
\hline
Normal & 0.003037838 & 0.05380121 & 0.05511659\\
Skew Normal & 0.003065280 & 0.05395187 & 0.05536497\\
t-student & 0.003672891*** & 0.05874602*** & 0.06060438***\\
Skew t-student & 0.003677610*** & 0.05883103*** & 0.06064330***\\
GED & 0.003183758*** & 0.05485659*** & 0.05642480***\\
Skew GED & 0.003285759*** & 0.05563557*** & 0.05732154***\\
\hline
\end{tabular}
\label{t17}
\begin{tablenotes}
      \small
      \item Note: *** means significance at 1\%, ** at 5\% and * at 10\%, otherwise no significance for \cite{diebold2002comparing} test of predictive accuracy compared with GARCH(1,1) under normal distribution. Under the null we have equal predictive accuracy.
\end{tablenotes}
\end{table}

In evaluating the forecasting performances, the best model is the GARCH(1,1) one based on Skew Normal distribution, even if according to the alternative loss functions MAE and RMSE the differences in predictive accuracy with respect to the Gaussian GARCH(1,1) are not statistically significant.

According to \cite{diebold2002comparing} test of predictive accuracy, most of the models statistically differ and outperform the selected benchmark.

Comparing, instead, predictive accuracy between skewed and not skewed models, we found that the Gaussian distribution is not statistically different in most of cases from its skewed extension. The same applies for GED. For $t$-student distribution family, this indifference applies three times (see Table \ref{t18}).

\begin{table}[H]
\centering
\caption{Predictive accuracy test between skewed and not skewed models}
\begin{tabular}{l l l l}
\hline
\textbf {Distributions} & {} & {}\\
\hline
\textit{GARCH(1,1)} & {Skewed Normal} & {Skewed t-student} & {Skewed GED}\\
\hline
Normal & 1.842 & -19.264*** & -9.2617*** \\
t-student &  17.904*** & -4.2316*** &  18.118*** \\
GED & 8.7561*** & -23.986*** & -2.2306**\\
\hline
\textit{GJR-GARCH(1,1)} & {Skewed Normal} & {Skewed t-student} & {Skewed GED}\\
\hline
Normal & -0.66937 & -16.009*** & -8.8147*** \\
t-student & 15.909*** & -3.01*** &  18.939*** \\
GED & 8.9696*** & -19.772*** & 1.2921\\
\hline
\textit{T-GARCH(1,1)} & {Skewed Normal} & {Skewed t-student} & {Skewed GED}\\
\hline
Normal & 1.3294 & -18.205*** & -7.587*** \\
t-student & 17.53*** & -4.3987*** &  17.797*** \\
GED & 9.1048*** &  -23.623*** & -0.54245 \\
\hline
\textit{E-GARCH(1,1)} & {Skewed Normal} & {Skewed t-student} & {Skewed GED} \\
\hline
Normal & 4.0673*** & -17.724*** & -7.9424*** \\
t-student & 17.875*** & 0.26758 &  19.978*** \\
GED & 9.1048*** & -15.898*** & 6.0024***\\
\hline
\textit{I-GARCH(1,1)} & {Skewed Normal} & {Skewed t-student} & {Skewed GED}\\
\hline
Normal & 3.5548*** & -18.606*** & -3.5098*** \\
t-student & 17.23*** & -4.3958 &  19.325*** \\
GED & 4.1211*** & -28.818*** & -0.3609***\\
\hline
\textit{AP-ARCH(1,1)} & {Skewed Normal} & {Skewed t-student} & {Skewed GED}\\
\hline
Normal & -1.7257* & -16.211 & -8.2565 \\
t-student & 13.938*** & -0.87949 &  11.925*** \\
GED & 4.2151** & -20.184*** & -5.3566***\\
\hline
\end{tabular}
\label{t18}
\begin{tablenotes}
      \small
      \item Note: the reported values are associated to the results of \cite{diebold2002comparing} test statistic under MSE loss function. *** means significance at 1\%, ** at 5\% and * at 10\%, otherwise no significance. Under the null we have equal predictive accuracy.
\end{tablenotes}
\end{table}

Nevertheless, we can recognize significant differences between alternative distribution families. In this sense, the predictive accuracy test reveals statistically different forecasts between $t$-student versus Generalized Error Distribution, as well as differences between their skewed extensions.

In conclusion, even if in this experiment Skewed GED is not the distributional assumption related to most performing model for both in-sample -- for which it represents the best assumption -- and out-of-sample analysis -- where the skewed normal distribution is as the best one--, it is surely the best alternative in capturing heavy-tails and skewness in returns.

\subsection{Litecoin data}
\label{sublit}

The last experiment is conducted on a cryptocurrency with a lower market capitalization. Indeed, Litecoin is the fifth ranked cryptocurrency in terms of market capitalization. Nevertheless, also Litecoin is also one of the cryptocurrencies with the highest volumes (https://coinmarketcap.com). As in the first application, we study the dynamics of the exchange rate with US Dollars.

As for the other two experiments, the first step is to assess for the model specification and parameters estimation, proving that data are not normally distributed.

The result of the Jarque-Bera test is 10861.76 with a p-value close to zero, which means that we can reject the null hypothesis that residuals follow a normal distribution also in this case. So, these results allow us to specify a GARCH model based on alternative distributions instead of a Gaussian-type GARCH model.

Then, proceeding with the parameters estimation of the standard GARCH(1,1) model based on normality, we found the results for variance equation represented in Table \ref{t19}.

\begin{table}[H]
\centering
\caption{Estimation for Gaussian GARCH(1,1) model}
\begin{tabular}{l l l}
\hline
\textbf{} & {Coefficient} & {Standard Error}\\
\hline
$\omega$ & 0.000091* & 0.000050 \\
$\alpha$ & 0.061723*** & 0.017813 \\
$\beta$ & 0.916084*** & 0.016773 \\
\hline
\end{tabular}
\begin{tablenotes}
      \small
      \item Note: *** means significance at 1\%, ** at 5\% and * at 10\%, standard errors are computed as robust.
    \end{tablenotes}
\label{t19}
\end{table}

\begin{figure}[H]
    \centering
    \includegraphics[width=\textwidth]{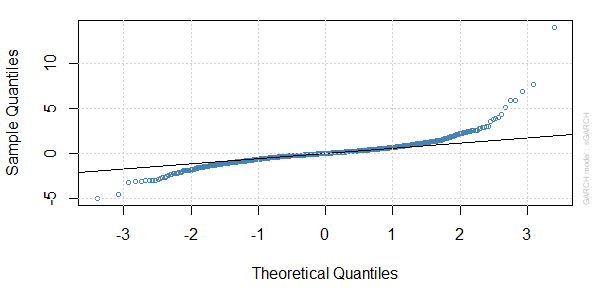}
    \caption{Q-Q plot of standardized residuals from Gaussian GARCH(1,1).}
\label{Fig7}
\end{figure}

After the estimation of the parameters, we have analyzed also the Q-Q plot of standardized residuals to test if normality assumption holds for the specified model (see Figure \ref{Fig7}).

Considering the residuals shape in the plot, the normality assumption seems again to be violated. Therefore we can estimate alternative GARCH(1,1) models for the volatility.

In so doing, we recognize the GARCH(1,1) model under Skewed GED assumption as the one with the best estimates. The results are shown in the Table \ref{t20}.

\begin{table}[H]
\centering
\caption{Results from the alternative GARCH(1,1) models}
\begin{tabular}{l l l l l l}
\hline
\textbf{} & {Skew Normal} & {t-student} & {Skew t-student} & {GED} & {Skew GED}\\
\hline
$\omega$ & 0.000062 & 0.000009 & 0.000009 & 0.000019 & 0.000016**\\
       & (0.000051)	 & (0.000007) & (0.000007) & (0.000014) & (0.000005)\\
$\alpha$ & 0.065986 & 0.081975*** & 	0.081785*** & 0.080730*** & 0.078319***\\
       & (0.090440)	 & (0.012143) & (0.012529)	& (0.017486) & (0.002721)\\
$\beta$  & 0.920405*** & 0.917025*** & 0.917215*** & 0.918269*** & 0.920676***\\
       & (0.026676)	 & (0.016637) & (0.016865) & (0.018805) & (0.006666)\\
\hline
\end{tabular}
\begin{tablenotes}
      \small
      \item Note: *** means significance at 1\%, ** at 5\% and * at 10\%, robust standard errors in parenthesis.
    \end{tablenotes}
\label{t20}
\end{table}

The in-sample analysis has been also implemented (see Table \ref{t20}). The most noticeable result is that Gaussian GARCH models arise as the ones with worst fitting.

Among the wide class of considered models, the skewed distributions show the most accurate fitting in terms of the in-sample analysis (see Tables \ref{t21} and \ref{t22}). More precisely, the $t$-student family slightly outperforms the GED in this case.

\begin{table}[H]
\centering
\caption{Information criteria for all GARCH models}
\begin{tabular}{l l l}
\hline
\textbf {Distribution} & {AIC} & {BIC}\\
\hline
GARCH(1,1)\\
\hline
Normal & -3.0264 & -2.9942 \\
Skew Normal & -3.1300 & -3.0938 \\
t-student & -3.6336 & -3.5974 \\
Skew t-student & -3.6348 & -3.5945 \\
GED & -3.5979 & -3.5617 \\
Skew GED & -3.6059 & -3.6059 \\
\hline
GJR-GARCH(1,1)\\
\hline
Normal & -3.0740 & -3.0378 \\
Skew Normal & -3.1063 & -3.0661 \\
t-student & -3.6381 & -3.5979 \\
Skew t-student & -3.6397 & -3.5954 \\
GED &  -3.6007 & -3.5604 \\
Skew GED & -3.6091 & -3.5648 \\
\hline
T-GARCH(1,1)\\
\hline
Normal & -3.0904 & -3.0541 \\
Skew Normal & -3.1395 & -3.0993 \\
t-student & -3.6523 & -3.6121 \\
Skew t-student & -3.6551 & -3.6108 \\
GED & -3.2298 & -3.1895 \\
Skew GED & -3.2282 & -3.1840 \\
\hline
\end{tabular}
\label{t21}
\begin{tablenotes}
      \small
      \item Note: AIC and BIC are Akaike Information Criterion and Bayesian Information Criterion, respectively. The lowest value is associated to the best fitting.
\end{tablenotes}
\end{table}

\begin{table}[H]
\centering
\caption{Information criteria for all GARCH models}
\begin{tabular}{l l l}
\hline
\textbf {Distribution} & {AIC} & {BIC}\\
\hline
E-GARCH(1,1)\\
\hline
Normal & -3.0848 & -3.0485 \\
Skew Normal & -3.1426 & -3.1023 \\
t-student & -3.6536 & -3.6134 \\
Skew t-student & -3.6550 & -3.6107 \\
GED & -3.6094 & -3.6094 \\
Skew GED & -3.6159 & -3.5716 \\
\hline
I-GARCH(1,1)\\
\hline
Normal & -3.0150 & -2.9869 \\
Skew Normal & -3.0836 & -3.0514 \\
t-student & -3.6356 & -3.6034 \\
Skew t-student & -3.6368 & -3.6006 \\
GED & -3.5987 & -3.5665 \\
Skew GED & -3.6077 & -3.5714 \\
\hline
AP-ARCH(1,1)\\
\hline
Normal & -3.0968 & -3.0566 \\
Skew Normal & -3.1063 & -3.0620 \\
t-student & -3.6512 & -3.6069 \\
Skew t-student & -3.6533 & -3.6050 \\
GED & -3.6031 & -3.5589 \\
Skew GED & -3.6174 & -3.5691 \\
\hline
\end{tabular}
\label{t22}
\begin{tablenotes}
      \small
      \item Note: AIC and BIC are Akaike Information Criterion and Bayesian Information Criterion, respectively. The lowest value is associated to the best fitting.
\end{tablenotes}
\end{table}

In the out-of-sample analysis we evaluate the forecasting accuracy of the models. The resulting quality of the forecasts is presented in Tables \ref{t23} and \ref{t24}.

\begin{table}[H]
\centering
\caption{Volatility forecasting performance for GARCH(1,1)-type models}
\begin{tabular}{l l l l}
\hline
\textbf {Distribution} & {MSE} & {MAE} & {RMSE}\\
\hline
GARCH(1,1)\\
\hline
Normal$^{\dagger}$ & 0.00282721 & 0.05242887 &  0.05317154\\
Skew Normal &  0.00296847*** & 0.05335918*** & 0.05448375***\\
t-student & 0.00258173*** & 0.04935203*** & 0.05081074*** \\
Skew t-student & 0.00258301*** & 0.04937885*** & 0.05082340*** \\
GED & 0.00268209*** & 0.05038674*** &  0.05178890***\\
Skew GED & 0.00266673*** & 0.05030135*** & 0.05164048***\\
\hline
GJR-GARCH(1,1)\\
\hline
Normal & 0.00696783*** & 0.07863949*** & 0.08347354***\\
Skew Normal & 0.00246737*** & 0.04811665*** & 0.04967271*** \\a
t-student & 0.00249264*** & 0.04807516*** & 0.04992636***\\
Skew t-student & 0.00251334*** & 0.04828849*** & 0.05013326*** \\
GED &  0.00257018*** & 0.04883740*** &  0.05069703***\\
Skew GED & 0.00255713*** & 0.04882139*** & 0.05056816***\\
\hline
T-GARCH(1,1)\\
\hline
Normal & 0.00267514** & 0.04984761*** &   0.05172182***\\
Skew Normal & 0.00261596*** & 0.04940308*** & 0.05114654***\\
t-student & 0.00518997*** & 0.06872743*** & 0.07204149*** \\
Skew t-student & 0.00514140*** & 0.06845046*** & 0.07170361*** \\
GED & 0.00032376*** & 0.01690694*** & 0.01799348***\\
Skew GED & 0.00032376*** & 0.01690694*** &  0.01799348***\\
\hline
\end{tabular}
\label{t23}
\begin{tablenotes}
      \small
      \item Note: *** means significance at 1\%, ** at 5\% and * at 10\%, otherwise no significance for \cite{diebold2002comparing} test of predictive accuracy compared with GARCH(1,1) under normal distribution ($\dagger$ recognizes the benchmark model). Under the null we have equal predictive accuracy.
\end{tablenotes}
\end{table}

\begin{table}[H]
\centering
\caption{Volatility forecasting performance for GARCH(1,1)-type models}
\begin{tabular}{l l l l}
\hline
\textbf {Distribution} & {MSE} & {MAE} & {RMSE}\\
\hline
E-GARCH(1,1)\\
\hline
Normal & 0.00267379** &  0.04984862*** & 0.05170876*** \\
Skew Normal & 0.00267408** & 0.04998585*** & 0.05171157***\\
t-student & 0.00554498*** & 0.07152296*** & 0.07446465***\\
Skew t-student & 0.00541826*** & 0.07078403*** & 0.07360889***\\
GED & 0.00281408 & 0.05095062 & 0.05304798\\
Skew GED & 0.00278774 & 0.05082947 & 0.05279906\\
\hline
I-GARCH(1,1)\\
\hline
Normal & 0.00335146*** & 0.05686214*** & 0.05789184*** \\
Skew Normal & 0.00323004*** & 0.05574411*** & 0.05683350*** \\
t-student & 0.0026129*** & 0.04965375*** & 0.05111710***\\
Skew t-student & 0.00261471*** & 0.04968532*** & 0.05113428***\\
GED & 0.00267455*** & 0.05031714*** &  0.05171611***\\
Skew GED &  0.00270490*** & 0.05062640*** & 0.05200872*** \\
\hline
AP-ARCH(1,1)\\
\hline
Normal & 0.00278435 & 0.05031387 & 0.05276699\\
Skew Normal & 0.00280012 & 0.05124268 & 0.05291620\\
t-student & 0.00488870*** & 0.06695473*** &  0.06991931***\\
Skew t-student & 0.00488501*** & 0.06693791*** & 0.06989290***\\
GED & 0.00271051 & 0.04950270 & 0.05206259\\
Skew GED & 0.00282363 & 0.05086309 & 0.05313789\\
\hline
\end{tabular}
\label{t24}
\begin{tablenotes}
      \small
      \item Note: *** means significance at 1\%, ** at 5\% and * at 10\%, otherwise no significance for \cite{diebold2002comparing} test of predictive accuracy compared with GARCH(1,1) under normal distribution. Under the null we have equal predictive accuracy.
\end{tablenotes}
\end{table}

For Litecoin data, the evaluation of the forecasting performance allows us to identify the best distribution assumption as the Skewed GED, even if -- as we already said above -- the in-sample analysis provides slightly different results. This finding is in line with the one related to Bitcoin data. Therefore, still a skewed model guarantees better forecasting performances.

In the end, we provide an evaluation of difference in predictive accuracy between skewed and not skewed models for all the alternatives GARCH(1,1)-type specifications (Table \ref{t25}).

\begin{table}[H]
\centering
\caption{Predictive accuracy test between skewed and not skewed models}
\begin{tabular}{l l l l}
\hline
\textbf {Distributions} & {} & {}\\
\hline
\textit{GARCH(1,1)} & {Skewed Normal} & {Skewed t-student} & {Skewed GED}\\
\hline
Normal & -4.296*** & 9.9645*** & 7.0772*** \\
t-student & -12.58*** & -0.68621 &  -16.489*** \\
GED & -9.0973*** & 35.158*** & 2.7805***\\
\hline
\textit{GJR-GARCH(1,1)} & {Skewed Normal} & {Skewed t-student} & {Skewed GED}\\
\hline
Normal & 16.335*** & 16.456*** & 16.182*** \\
t-student & 0.89782 & -24.887*** &  -26.619*** \\
GED & 4.1846*** & 14.032*** & 2.5137**\\
\hline
\textit{T-GARCH(1,1)} & {Skewed Normal} & {Skewed t-student} & {Skewed GED}\\
\hline
Normal & 6.4152*** & -18.681*** & 24.202*** \\
t-student & 18.674*** & 9.6102*** &  23.14*** \\
GED & -25.389*** & -23.225*** & 0 \\
\hline
\textit{E-GARCH(1,1)} & {Skewed Normal} & {Skewed t-student} & {Skewed GED} \\
\hline
Normal & 0.086513 & -21.826*** & -2.829*** \\
t-student & 22.024*** & 15.192*** &  24.175*** \\
GED & 4.7704*** & -25.508*** & 4.6938***\\
\hline
\textit{I-GARCH(1,1)} & {Skewed Normal} & {Skewed t-student} & {Skewed GED}\\
\hline
Normal & 16.24*** & 73.784*** & 58.138*** \\
t-student & -54.1*** & -1.01 &  -23.037*** \\
GED & -44.075*** & 19.634*** & -6.7331***\\
\hline
\textit{AP-ARCH(1,1)} & {Skewed Normal} & {Skewed t-student} & {Skewed GED}\\
\hline
Normal & -0.57657 & -18.925*** & -1.0195 \\
t-student & 17.145*** & 1.0763 &  22.527*** \\
GED & -2.5253** & -21.708*** & -5.1806***\\
\hline
\end{tabular}
\label{t25}
\begin{tablenotes}
      \small
      \item Note: the reported values are associated to the results of \cite{diebold2002comparing} test statistic under MSE loss function. *** means significance at 1\%, ** at 5\% and * at 10\%, otherwise no significance. Under the null we have equal predictive accuracy.
\end{tablenotes}
\end{table}

According to this experiment, the Skew $t$-student distribution fails to provide statistically different forecasts compared to its symmetric version, while for the other two families of distributions -- i.e., Gaussian and GED -- the converse situation applies.

Indeed, we can argue that Skewed Normal/GED statistically outperforms the standard Gaussian/GED. Moreover, the Skewed GED provides the best forecast accuracy among all the other alternatives.

\section{Further robustness checks}
\label{secrobu}

In this section we provide evidence of robustness about the results presented in the previous Section. We consider first changes in forecasting scheme and testing set.

Previous results are based on rolling window scheme; therefore, we here present robustness according to a recursive scheme.

Moreover, we provide also evidence of robustness of the obtained findings by changing the length of the testing set.

Then, in the last subsection, we present alternative forecasting models of non GARCH-type and apply them for volatility prediction purposes. In so doing, we give further support to our methological proposal. Indeed, as we will see below, all the considered models underperform the best one we found within the GARCH-type framework, in all the analyzed cases of exchange rates between cryptocurrencies and USD.

\subsection{Forecasting with recursive approach}

The idea of the recursive approach is quite similar to the rolling window, with a remarkable distinction. Indeed, in the recursive approach we firstly consider the initial time-window with 200 time data. Then, such a window is moved by including one-day ahead. However, in the recursive approach here employed, the first day is not excluded, so that the time-window is enlarged by one unit at each recursive time step.

As robustness check, we evaluate the out-of-sample performances of all the considered volatility models according to the recursive scheme. The results related to Bitcoin/USD exchange rate are showed in the Table \ref{t26}, while for Ethereum/USD and Litecoin/USD results are in the Table \ref{t27} and Table \ref{t28}, respectively.
\begin{table}[H]
\centering
\caption{Forecasting accuracy with recursive approach: Bitcoin/USD}
\begin{tabular}{l l l l}
\hline
\textbf {} & {GARCH} & {GJR-GARCH} & {T-GARCH}\\
\hline
Normal & 0.00226652$^\dagger$ & 0.00231733*** & 0.00360885*** \\
t-student & 0.00348222*** & 0.00331076*** & 0.00309894*** \\
GED & 0.00336300*** & 0.00324299*** & 0.00001424***\\
Skew Normal & 0.00224931*** & 0.00225703*** & 0.00346474***\\
Skew t-student & 0.00350264*** & 0.00332987*** & 0.00311676***\\
Skew GED & 0.00336217*** & 0.00324666*** & 0.00001424***\\
\hline
\textbf {} & {E-GARCH} & {I-GARCH} & {AP-ARCH}\\
\hline
Normal & 0.00235180*** & 0.00525533*** & 0.00340147*** \\
t-student & 0.00178647*** & 0.00374075*** &  0.00264207*** \\
GED & 0.00116924*** & 0.00360469*** & 0.00155773***\\
Skew Normal & 0.00230088*** & 0.00530999*** & 0.00273173***\\
Skew t-student & 0.00177781*** & 0.00376253*** & 0.00247391***\\
Skew GED & 0.00116312*** & 0.00360491*** & 0.00235783***\\
\hline
\end{tabular}
\label{t26}
\begin{tablenotes}
      \small
      \item Note: the reported values are associated to the MSE loss function. *** means significance at 1\%, ** at 5\% and * at 10\%, otherwise no significance for \cite{diebold2002comparing} test of predictive accuracy compared with GARCH(1,1) under normal distribution (highlighted with $\dagger$ symbol). Under the null we have equal predictive accuracy.
\end{tablenotes}
\end{table}

\begin{table}[H]
\centering
\caption{Forecasting accuracy with recursive approach: Ethereum/USD}
\begin{tabular}{l l l l}
\hline
\textbf {} & {GARCH} & {GJR-GARCH} & {T-GARCH}\\
\hline
Normal & 0.0040532$^\dagger$ & 0.00403640*** & 0.0045778*** \\
t-student & 0.0240013*** & 0.0244118*** & 0.0049275*** \\
GED & 0.0058418*** & 0.0069362*** & 0.0035572***\\
Skew Normal &  0.0040691*** & 0.0040638*** & 0.0045557***\\
Skew t-student & 0.0234560*** & 0.0237631*** &  0.0048822***\\
Skew GED & 0.0059137*** & 0.0071127*** & 0.0035495***\\
\hline
\textbf {} & {E-GARCH} & {I-GARCH} & {AP-ARCH}\\
\hline
Normal & 0.0040168*** & 0.0178729*** & 0.0040347 \\
t-student & 0.0039993*** & 0.0257453*** &  0.0065273*** \\
GED & 0.0032169*** & 0.0251742*** & 0.0045723***\\
Skew Normal & 0.0040224*** & 0.0175863*** & 0.0040499***\\
Skew t-student & 0.0039608*** & 0.00376253*** & 0.0061513***\\
Skew GED & 0.0031186*** & 0.0206809*** & 0.0059800***\\
\hline
\end{tabular}
\label{t27}
\begin{tablenotes}
      \small
      \item Note: the reported values are associated to the MSE loss function. *** means significance at 1\%, ** at 5\% and * at 10\%, otherwise no significance for \cite{diebold2002comparing} test of predictive accuracy compared with GARCH(1,1) under normal distribution (highlighted with $\dagger$ symbol). Under the null we have equal predictive accuracy.
\end{tablenotes}
\end{table}

\begin{table}[H]
\centering
\caption{Forecasting accuracy with recursive approach: Litecoin/USD}
\begin{tabular}{l l l l}
\hline
\textbf {} & {GARCH} & {GJR-GARCH} & {T-GARCH}\\
\hline
Normal & 0.0040532$^\dagger$ & 0.0040364*** & 0.0045778*** \\
t-student & 0.0240013*** & 0.0244118***  & 0.0049275*** \\
GED & 0.0058418*** & 0.0069362*** & 0.0035572***\\
Skew Normal &  0.0040691*** & 0.0040638*** & 0.0045557***\\
Skew t-student & 0.0234561*** & 0.0237631*** & 0.0048822***\\
Skew GED & 0.0059137*** & 0.0071127*** & 0.0035495***\\
\hline
\textbf {} & {E-GARCH} & {I-GARCH} & {AP-ARCH}\\
\hline
Normal &  0.0040168*** & 0.0178729*** & 0.0040347* \\
t-student & 0.0039993*** & 0.0257453*** &  0.0065273*** \\
GED & 0.0032169*** & 0.0206809*** & 0.0045724***\\
Skew Normal & 0.0040224*** & 0.0175863*** & 0.0061513***\\
Skew t-student & 0.0039608*** & 0.0251742*** & 0.00247391***\\
Skew GED & 0.0031186*** & 0.0202550*** &  0.0021187***\\
\hline
\end{tabular}
\label{t28}
\begin{tablenotes}
      \small
      \item Note: the reported values are associated to the MSE loss function. *** means significance at 1\%, ** at 5\% and * at 10\%, otherwise no significance for \cite{diebold2002comparing} test of predictive accuracy compared with GARCH(1,1) under normal distribution (highlighted with $\dagger$ symbol). Under the null we have equal predictive accuracy.
\end{tablenotes}
\end{table}

As clearly shown in all the tables above, prediction accuracy results are not affected by the employed type of forecasting scheme. In particular, in the case of Bitcoin/USD exchange rate, the T-GARCH based on GED and Skewed GED distributions significantly outperform all the alternatives. The same conclusions apply to the Litecoin/USD exchange rate.

A difference can be noted in the case related to the Ethereum/USD exchange rate. Indeed, as highlited in Section \ref{sec5.2}, the best distribution assumption has been proven to be the GARCH under Skew Normal distribution, still reflecting the relevance of skewness in the volatility models for crypotcurrencies.

However, according to the results obtained by implementing the recursive approach, there is a clear evidence of overperformance for the E-GARCH model under Skewed GED distribution assumption.

This said, the Skewed GED is confirmed to be the best assumption for all the considered cryptocurrencies, as already stated in the rolling window case presented in Section \ref{sec2}. Therefore, we get still stronger evidence in favor of the statistical model presented in the original analysis.

Notice that all the results shown Tables \ref{t26}, \ref{t27} and \ref{t28} are related to the Mean Square Error, that is the most robust loss function according to \cite{patton2011volatility}. However, results actually hold also for the other considered loss functions, as unreported tables highlight.

\subsection{Forecasting with a different testing set}
In this case, we implement a forecast exercise on a rolling windows basis, by taking 100 units of time as testing set, instead of the 200 ones employed in the original analysis.

\begin{table}[H]
\centering
\caption{Forecasting accuracy with testing set as last 100 observations: Bitcoin/USD}
\begin{tabular}{l l l l}
\hline
\textbf {} & {GARCH} & {GJR-GARCH} & {T-GARCH}\\
\hline
Normal & 0.00166181$^\dagger$ &  0.00171153*** & 0.00180141*** \\
t-student & 0.00168049*** & 0.00164767*** & 0.00271630*** \\
GED & 0.00168347*** &  0.00165776*** & 0.00019423***\\
Skew Normal & 0.00165741*** & 0.00167630*** & 0.00175730***\\
Skew t-student & 0.00167772*** & 0.00164494*** & 0.00273767***\\
Skew GED & 0.00168254*** & 0.00165576*** & 0.00019423***\\
\hline
\textbf {} & {E-GARCH} & {I-GARCH} & {AP-ARCH}\\
\hline
Normal & 0.00175046*** & 0.00187223*** & 0.00175829*** \\
t-student & 0.00256342*** & 0.00168988*** &  0.00271662*** \\
GED & 0.00168573*** & 0.00169242*** &  0.00161437***\\
Skew Normal &  0.00170528*** & 0.00184210*** & 0.00271895***\\
Skew t-student & 0.00256322*** & 0.00168706*** & 0.00247391***\\
Skew GED & 0.00169123*** & 0.00169756*** & 0.00178099***\\
\hline
\end{tabular}
\label{t29}
\begin{tablenotes}
      \small
      \item Note: the reported values are associated to the MSE loss function. *** means significance at 1\%, ** at 5\% and * at 10\%, otherwise no significance for \cite{diebold2002comparing} test of predictive accuracy compared with GARCH(1,1) under normal distribution (highlighted with $\dagger$ symbol). Under the null we have equal predictive accuracy.
\end{tablenotes}
\end{table}

\begin{table}[H]
\centering
\caption{Forecasting accuracy with testing set as last 100 observations: Ethereum/USD}
\begin{tabular}{l l l l}
\hline
\textbf {} & {GARCH} & {GJR-GARCH} & {T-GARCH}\\
\hline
Normal & 0.0016618$^\dagger$ &  0.0033844*** & 0.0033998*** \\
t-student & 0.0042598*** & 0.0042491*** & 0.0051203*** \\
GED & 0.0037651*** &  0.0037023*** & 0.0038606***\\
Skew Normal & 0.0034106*** & 0.0033680*** & 0.0036183***\\
Skew t-student & 0.0042753*** & 0.0042626*** & 0.0051229***\\
Skew GED & 0.0036715*** & 0.0037174*** & 0.0037979***\\
\hline
\textbf {} & {E-GARCH} & {I-GARCH} & {AP-ARCH}\\
\hline
Normal & 0.0033932*** & 0.0040476*** & 0.00337533*** \\
t-student & 0.0049396*** & 0.0042743*** &  0.0050415*** \\
GED & 0.0036938*** & 0.0040601*** &  0.0037595***\\
Skew Normal &  0.0033940*** & 0.0040507*** & 0.0033964***\\
Skew t-student & 0.0049336*** & 0.0042903*** & 0.0050622***\\
Skew GED & 0.0037406*** & 0.0040480*** & 0.0037519***\\
\hline
\end{tabular}
\label{t30}
\begin{tablenotes}
      \small
      \item Note: the reported values are associated to the MSE loss function. *** means significance at 1\%, ** at 5\% and * at 10\%, otherwise no significance for \cite{diebold2002comparing} test of predictive accuracy compared with GARCH(1,1) under normal distribution (highlighted with $\dagger$ symbol). Under the null we have equal predictive accuracy.
\end{tablenotes}
\end{table}

\begin{table}[H]
\centering
\caption{Forecasting accuracy with testing set as last 100 observations: Litecoin/USD}
\begin{tabular}{l l l l}
\hline
\textbf {} & {GARCH} & {GJR-GARCH} & {T-GARCH}\\
\hline
Normal & 0.00355692$^\dagger$ &  0.00336571*** & 0.00359048*** \\
t-student & 0.00351696*** & 0.00354558*** & 0.00820642*** \\
GED & 0.00354603*** &  0.00365435*** &  0.00047286***\\
Skew Normal & 0.00355154*** & 0.00337813*** & 0.00356620***\\
Skew t-student & 0.00351464*** & 0.00356696*** & 0.00822959***\\
Skew GED & 0.00358403*** & 0.00356236*** &  0.00047286***\\
\hline
\textbf {} & {E-GARCH} & {I-GARCH} & {AP-ARCH}\\
\hline
Normal & 0.00362372*** & 0.00420821*** & 0.00476177*** \\
t-student & 0.00882732*** & 0.00355872*** & 0.00748153*** \\
GED & 0.00416083*** & 0.00365541*** &  0.00438798***\\
Skew Normal &  0.00361398*** & 0.00413231*** & 0.0033964***\\
Skew t-student & 0.00851569*** & 0.00355886*** & 0.00760975***\\
Skew GED & 0.00407516*** & 0.00365657*** & 0.00376108***\\
\hline
\end{tabular}
\label{t31}
\begin{tablenotes}
      \small
      \item Note: the reported values are associated to the MSE loss function. *** means significance at 1\%, ** at 5\% and * at 10\%, otherwise no significance for \cite{diebold2002comparing} test of predictive accuracy compared with GARCH(1,1) under normal distribution (highlighted with $\dagger$ symbol). Under the null we have equal predictive accuracy.
\end{tablenotes}
\end{table}

In the case of Bitcoin/USD exchange rate (Table \ref{t29}), results do not change with respect to those of the original analysis. Indeed, again we observe evidence in favor of the T-GARCH model under Skewed GED distribution. The same results apply to Litecoin/USD exchange rate (Table \ref{t31}), where we identify the T-GARCH model under Skewed GED distribution as the best one in terms of out-of-sample performance.
For the Ethereum/USD exchange rate (Table \ref{t30}) we do not obtain different results compared to the ones in Section \ref{sec5.2}, since the GARCH model under Skew Normal distribution still performs better in the out-of-sample exercise.

Hence, we get evidence of robustness also by changing the length of the testing set. Also in this case, the results shown in Tables \ref{t29}, \ref{t30} and \ref{t31} are related to the Mean Square Error. However, these results hold also for the other considered loss functions in unreported tables.

\subsection{Alternative forecasting volatility models}

In finance, the time-varying volatility of risky assets is usually modeled and predicted by using a number of GARCH-type models and their extensions; under this framework, the conditional variance of a risky asset is a deterministic function of model parameters and past data. The same argument applies also to cryptocurrencies, for which there is evidence of GARCH-type models for volatility (see the discussion in Section \ref{sec3} and the references therein quoted).

The results of our analysis (see Section \ref{secempi}) offer a not unique best model -- in terms of prediction performance of all the considered exchange rates between cryptocurrencies and USD -- for describing volatility. However, there is a clear evidence in favor of Skewed GED GARCH models.

In this section, as further robustness, we show that such results do not change also when the comparison analysis includes also a large number of models of non GARCH-type, i.e. GARCH models based on Skewed GED perform still better.
More specifically, among the other possibilities, we here deal with two of the most powerful tools for estimating volatility: Dynamic Score Models (DSC) and stochastic volatility models.


The standard stochastic volatility model can be defined as follows (see e.g. \cite{jacquier2002}):

\begin{equation}
\label{sv1}
    y_t = e^{h_t/2}\epsilon^{y_t},
\end{equation}
\begin{equation}
\label{ht}
    h_t = \mu_h + \phi_h (h_{t-1}-\mu_h) \epsilon^{h_t},
\end{equation}
\\
where both $\epsilon^{y_t}$ and $\epsilon^{h_t}$ are normally distributed, $|\phi_h| < 1$, $\mu_h>0$ and $h_t$ is the log-volatility. By (\ref{ht}), the log-volatility follows a Gaussian AR(1) process with conditional mean $\mu_h$. 
Since simulation efficiency in state-space models can often be improved through model reparametrizations, we follow the proposal of \cite{kastner2014} and the following parametrization of (\ref{sv1}--\ref{ht}):

\begin{equation}
y_t \sim N(0,\omega e^{h_t-\mu_h}),
\end{equation}
\begin{equation}
h_t-\mu_h  = \phi_h (h_{t-1}-\mu_h) \epsilon^h_t,
\end{equation}
where $\omega = e^{\mu_h}$. Then, we apply the algorithm proposed in \cite{kastner2014} to estimate the parameters, on the basis of an efficient Markov Chain Monte Carlo (MCMC) estimation scheme by specifying a Gaussian prior distribution. Then we use the MCMC algorithm to draw from the posterior distribution of the random variables in order to make forecasts. More specifically, we implement the MCMC sampler to obtain posterior draws given by $h_{1:t}$; then, we compute the predictive mean $E(h_{t+k}|h_{1:t})$. Next, we move one period ahead and repeat 1000 times the whole exercise with data $h_{1:t+1}$ and so forth, recursively.

In practice, the predictive mean of $h_{t+k}$ cannot be computed analytically. Instead, they are obtained by using predictive simulations. These forecasts are then averaged over all the posterior draws to produce estimates for $E(h_{t+k} | h_{1:t})$; then, the whole exercise is repeated by using data up to time $t + 1$ to produce $E(h_{t+k+1} |h_{1:t+1})$.

However, for robustness purposes, another relatively new class of volatility models is presented: the so-called Dynamic Conditional Score (DSC) models, introduced in \cite{creal2013}. The ground of this methodology lies in the fact that the GARCH models consider the squared demeaned returns as the drivers of time–variation in the conditional variance, independently from the shape of the conditional distribution of the return. Moving from this, \cite{creal2013} proposed to use the score of the conditional density function as the main driver of time–variation in the parameters of the time series process adopted for describing the data. Parameters in Dynamic Conditional Score models are easily estimated via Maximum Likelihood approach.

The general expression of the DCS model is given by:

\begin{equation}
\label{ft}
    f_t = \omega + \beta f_{t-1} + \alpha S_{t-1} \left[ \frac{\partial {\rm log} p(r_{t-1}|f_{t-1})}{\partial f_{t-1}} \right],
\end{equation}

where $f_t$ is a conditional time varying parameter (e.g. the volatility), $S_t$ is a score function, ${\rm log} p(r_{t-1}|f_{t-1})$ is the log probability density function. The main difference between the model (\ref{ft}) and the classical GARCH model in (\ref{garch}) can be found in the evolution of the volatility equation -- for the GARCH model, one has $f_t = \sigma^2_t$ -- which in (\ref{ft}) depends on the past values of the score of the conditional distribution instead of only on the squared returns. Moreover, the DCS model is more general than the GARCH one, since the score does not depend only on the second-order moments but on the overall probability distribution of the reference random variable. Yet, as in the GARCH case, it is possible to specify different densities to compute the conditional scores simply by changing the stochastic assumptions on ${\rm log} p(r_{t-1}|f_{t-1})$. Just to provide some examples, by assuming a t-student distribution or a skewed t-student we get a t-student DCS or a skewed-t DCS models (see e.g. \cite{harvey2014})

Volatility forecasting exercises under a DCS framework run as for the GARCH models; hence, the forecasting procedure is the same as the one described above in the paper.


In Tables \ref{t35}, \ref{t36} and \ref{t37} we show the results for all the cryptocurrencies in terms of forecasting accuracy of the following alternative models: stochastic volatility, Gaussian Dynamic Conditional Score (DCS) model, Skewed Normal DCS, t-student DCS and skewed-t DCS.

\begin{table}[H]
\centering
\caption{Forecasting accuracy with alternative models: Bitcoin/USD}
\begin{tabular}{l l l l}
\hline
\textbf {} & \bf{Best model} & \bf{Stoch. vol.} & \bf{Gaussian-DCS}\\
\hline
MSE & 0.00115839 &  0.06781723*** & 0.03353264*** \\
MAE & 0.01008683 & 0.2567941*** & 0.1743166*** \\
RMSE & 0.03403522 &  0.2604174*** &  0.1831192***\\
\hline
\textbf & \bf{Skew Normal DCS} & \bf{t-student DCS} & \bf{Skewed t DCS}\\
\hline
MSE & 0.2173271*** &  0.1035493*** & 0.1783751*** \\
MAE & 0.375288*** & 0.284592*** & 0.4045632*** \\
RMSE & 0.4661836*** &   0.3217908*** &  0.4223447***\\
\hline
\end{tabular}
\label{t35}
\begin{tablenotes}
      \small
      \item Note: Best model is the best according to GARCH-type of Table \ref{t10}, while "stoch. vol." stays for "stochastic volatility" model. The reported values are associated to the MSE, MAE and RMSE loss functions. *** means significance at 1\%, ** at 5\% and * at 10\%, otherwise no significance for \cite{diebold2002comparing} test of predictive accuracy compared with the best GARCH-type model. Under the null we have equal predictive accuracy.
\end{tablenotes}
\end{table}

According to Table \ref{t10}, for Bitcoin/USD exchange rate the best model is the Skew GED-GARCH(1,1) -- which is the reported best model in Table \ref{t35}. Particularly, as anticipated before, all the models here underperform the best we found within the GARCH-types. Among all the alternatives, the best two models are the stochastic volatility model and the Gaussian Dynamic Score one. Nevertheless, the model with the highest out-of-sample accuracy is still the T-GARCH(1,1) based on Skew GED.

\begin{table}[H]
\centering
\caption{Forecasting accuracy with alternative models: Ethereum/USD}
\begin{tabular}{l l l l}
\hline
\textbf {} & \bf{Best model} & \bf{Stoch. vol.} & \bf{Gaussian-DCS}\\
\hline
MSE & 0.00289673 &  0.2377882*** & 0.09396339*** \\
MAE &  0.0524247 & 0.4475882*** & 0.2869582*** \\
RMSE & 0.0538212 &  0.4876353*** &  0.3065345***\\
\hline
\textbf & \bf{Skew Normal DCS} & \bf{t-student DCS} & \bf{Skewed t DCS}\\
\hline
MSE & 0.2467824*** &  0.1101726*** & 0.07369849*** \\
MAE & 0.4090161*** & 0.3048068*** & 0.2666203*** \\
RMSE & 0.496772*** &  0.3319226*** &  0.2714747***\\
\hline
\end{tabular}
\label{t36}
\begin{tablenotes}
      \small
      \item Note: Best model is the best according to GARCH-type Table \ref{t16}, while "stoch. vol." stays for "stochastic volatility" model. The reported values are associated to the MSE, MAE and RMSE loss functions. *** means significance at 1\%, ** at 5\% and * at 10\%, otherwise no significance for \cite{diebold2002comparing} test of predictive accuracy compared with the best GARCH-type model. Under the null we have equal predictive accuracy.
\end{tablenotes}
\end{table}

With respect to the Ethereum/USD exchange rate, the most accurate model is the Skew Normal T-GARCH(1,1) and it is reported in Table \ref{t36}. As for the case of Bitcoin, the best GARCH-type model overperforms all the alternatives. Differences are actually also very large in numerical terms. Notice that the skewed-t Dynamic Conditional Score model is the second best one, even if it is very far from the best GARCH-type of Table \ref{t16}.

\begin{table}[H]
\centering
\caption{Forecasting accuracy with alternative models: Litecoin/USD}
\begin{tabular}{l l l l}
\hline
\textbf {} & \bf{Best model} & \bf{Stoch. vol.} & \bf{Gaussian-DCS}\\
\hline
MSE & 0.00032376 &  0.3462538*** & 0.01095247*** \\
MAE & 0.0169069 & 0.5735445*** & 0.101634*** \\
RMSE & 0.0179934 &  0.5884334*** &  0.104654***\\
\hline
\textbf & \bf{Skew Normal DCS} & \bf{t-student DCS} & \bf{Skewed t DCS}\\
\hline
MSE & 0.2436648*** &  0.1566804*** & 0.2890927*** \\
MAE & 0.4057249*** & 0.3355313*** & 0.5081017*** \\
RMSE & 0.4936241*** &  0.3958288*** & 0.5376734***\\
\hline
\end{tabular}
\label{t37}
\begin{tablenotes}
      \small
      \item Note: Best model is the best according to GARCH-type in Table \ref{t23}, while "stoch. vol." stays for "stochastic volatility" model. The reported values are associated to the MSE, MAE and RMSE loss functions. *** means significance at 1\%, ** at 5\% and * at 10\%, otherwise no significance for \cite{diebold2002comparing} test of predictive accuracy compared with the best GARCH-type model. Under the null we have equal predictive accuracy.
\end{tablenotes}
\end{table}

For the Litecoin/USD exchange rate models compared in Table \ref{t23}, we highlight the overperformance of the Skew GED T-GARCH(1,1) and report it -- for comparison purposes -- as best model in Table \ref{t37}.

Also in this case, other additional models are not able to achieve out of sample performances higher  than the ones of the best GARCH-type model. Therefore, on the light of these results, we have a successful robustness check of the results presented in this paper. 

In conclusion, there is a clear evidence  that the GARCH-type extensions allowing with skewed and flexible distributions perform better than the Dynamic Conditional Score models and the stochastic volatility, in all the cases of considered exchange rates.

\section{Conclusions}
\label{secconc}

This paper merges together financial stylized facts, forecasting exercise, risk analysis, probability distributions theory and the analysis of the cryptocurrencies features.
We discuss the volatility forecasting of the exchange rates between the most popular cryptocurrencies and the US Dollar.

We follow a GARCH-based approach for the modelization of the volatility, which is totally in line with the main financial risk literature. However, we depart from the standard Gaussian assumption, in order to be more tailored on the financial reality of the evolution of the cryptocurrencies. We use a GED approach for modeling the stochastic source of the volatility. Such a choice is particularly reasonable, in that GED distributions are versatile and include several well-established random variables as subcases.  More than this, we include also the distributional properties of the cryptocurrencies, and employ at this aim the skewed versions of the GED distributions.

The empirical exercise illustrates the most suitable source of stochasticity for modeling purposes and for effective prediction exercises, tending specifically towards the skewed GED distribution.

The methodological procedures here presented are rather general and can be successfully adopted in other contexts of volatility estimation.
Moreover, the obtained findings are relevant for financial industries practitioners, such as data scientists in investment fund or banks, as well as traders that build intelligent systems for trading purposes.

However, it is important to point out two weaknesses of our approach. First, the theoretical proposal has been validated on a sample which is remarkably representative -- very relevant cryptocurrencies and the USD, the most important physical currency --  but it is not universal, in that it does not consider all the exchange rates. Second, no attention is paid to the interactions of the obtained results with possible macroeconomic shocks. In this respect, we are well aware that a shock in the economic system might modify the patterns of exchange rates and the consequent forecasting exercises. These points -- with a more specific focus on the second one -- seem to be of particular interest and merit a devoted study. For this reason, we have inserted them in our future research agenda.

\end{document}